\documentclass[12pt]{amsart}
\usepackage{amsmath,amssymb,amsfonts,amsthm,amstext,graphicx,color,enumerate}
\usepackage{hyperref,cleveref,url,mathtools}
\usepackage{tikz}
\usetikzlibrary{arrows}
\usepackage[numbers,sort&compress]{natbib}
\usepackage{graphbox,graphicx}
\usepackage{hyperref}

\usepackage{caption}
\usepackage{subcaption}
\usepackage[final]{pdfpages}
\usepackage[margin=3cm]{geometry}

\newtheorem*{thm*}{Theorem}

\crefname{thm}{Theorem}{Theorems}
\crefname{lemma}{Lemma}{Lemmas}
\crefname{prop}{Proposition}{Propositions}
\crefname{cor}{Corollary}{Corollaries}
\crefname{section}{Section}{Sections}
\crefname{figure}{Figure}{Figures}

\renewcommand{\i}{\mathbf{i}}
\renewcommand{\j}{\mathbf{j}}
\renewcommand{\k}{\mathbf{k}}

\makeatletter
\newcommand{\addresseshere}{%
  \enddoc@text\let\enddoc@text\relax
}
\makeatother

\begin{document}
\title[Spinor linkage]{The spinor linkage -- a mechanical implementation of the plate trick}
\author[Holroyd]{Alexander E.\ Holroyd}
\address{School of Mathematics, University of Bristol, U.K.}
\email{a.e.holroyd@bristol.ac.uk}
\date{1 July 2021}

\begin{abstract}
  The plate trick or belt trick is a striking physical demonstration of properties of the double cover of the three-dimensional rotation group by the sphere of unit quaternions or spinors.  The two ends of a flexible object are continuously rotated with respect to each other.  Surprisingly, the object can be manipulated so as to avoid accumulating twists.  We present a new mechanical linkage that implements this task. It consists of a sequence of rigid bodies connected by hinge joints, together with a purely mechanical control mechanism.  It has one degree of freedom, and the motion is generated by simply turning a handle.
\end{abstract}

\maketitle

\section{Introduction}

\begin{figure}\centering
\begin{tikzpicture}[>=triangle 60,align=center]
    \node[anchor=south west,inner sep=0] (image) at (0,0) {\includegraphics[width=.5\textwidth]{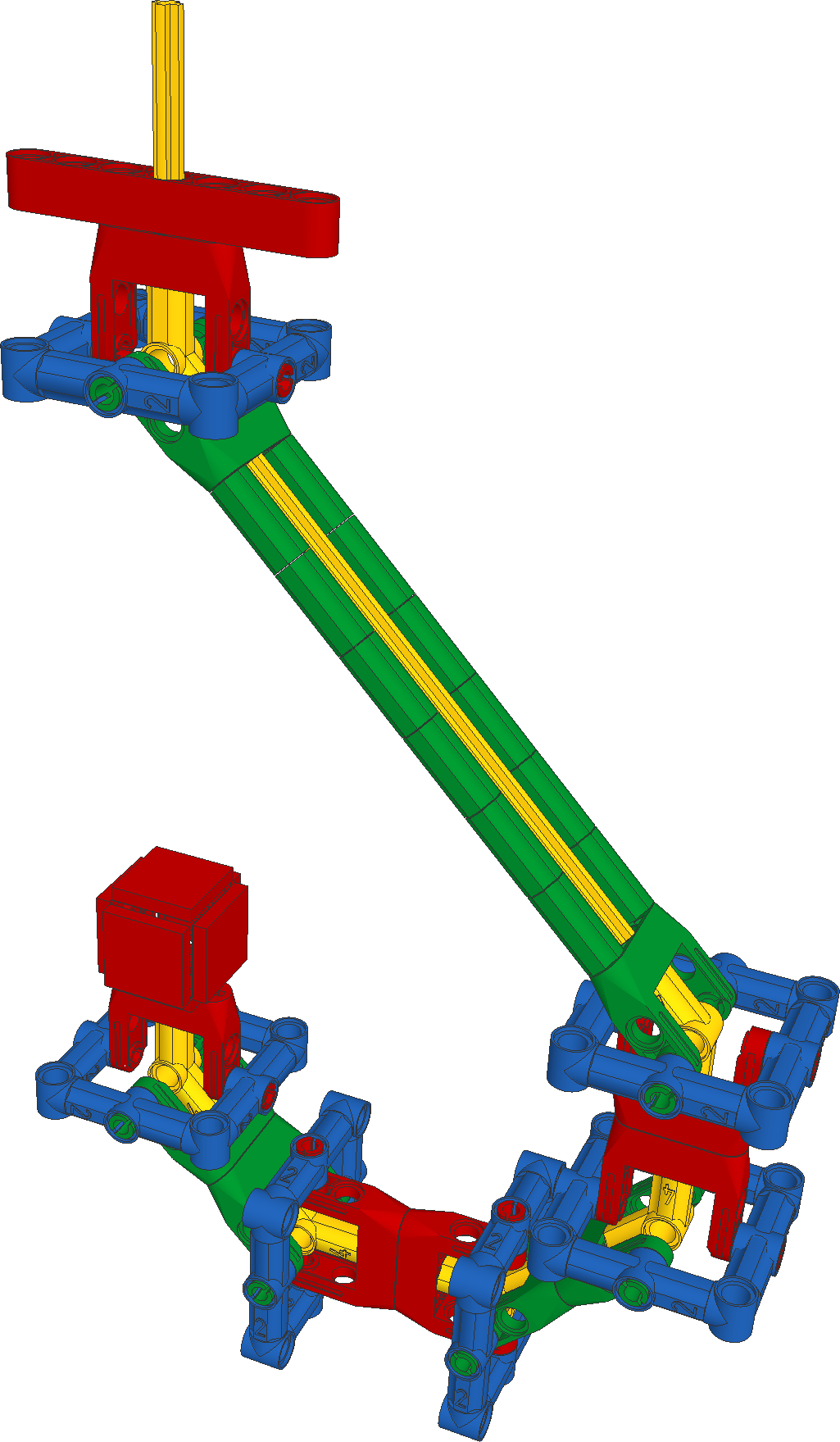}};
    \begin{scope}[
        x={(image.south east)},
        y={(image.north west)}]
        \draw[<-] (.05,.35) --+ (-.15,0) node[left]{spinning\\cube};
        \draw[<-] (.45,.85) --+ (.15,0) node[right]{fixed to\\frame};
        \draw[<-] (.45,.95) --+ (.15,0) node[right]{control\\shaft};
    \end{scope}
\end{tikzpicture}
  \caption{The Spinor Linkage.}
  \label{trunk}
\end{figure}
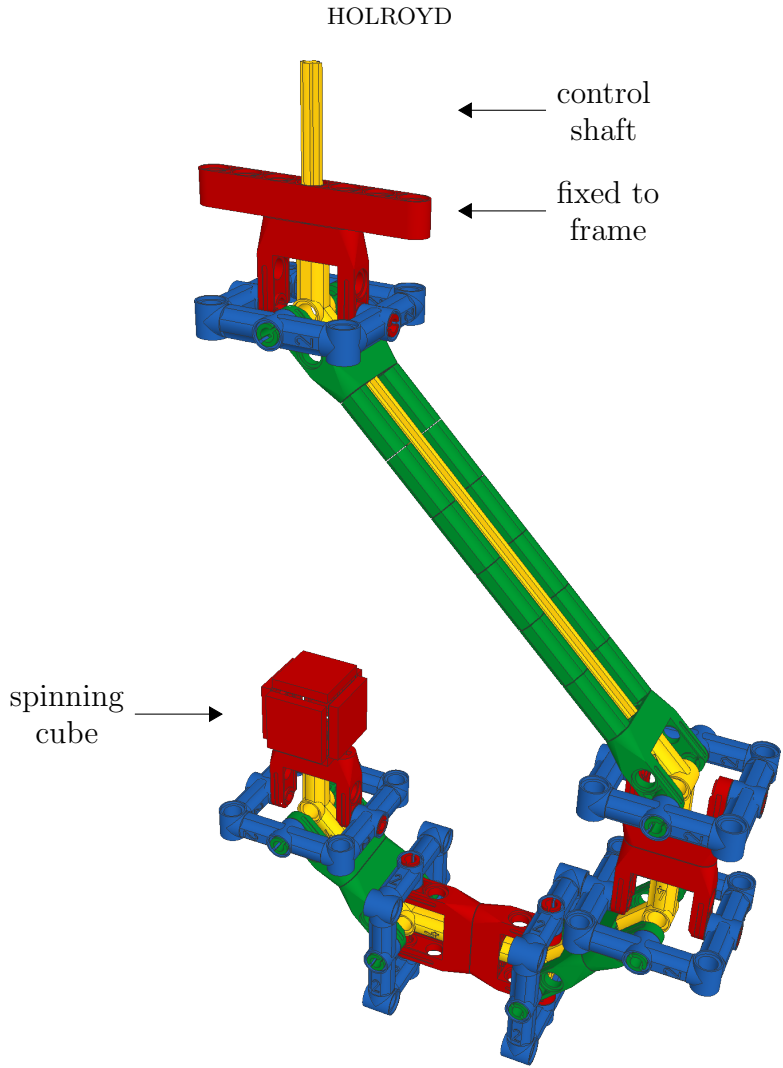

Consider a rigid object, spinning continuously at constant angular velocity around a fixed axis, say a vertical one.  Suppose that the object is tethered to the ground via a flexible connector, such as a rope, belt or cable, which is fixed rigidly to the object at one end, and to the ground at the other.  What will happen to the connector as the rotation continues?  Will it inevitably become more and more twisted (to the point of physical failure)?  Or is it possible to somehow untwist it after each full rotation?

The startling truth is that neither answer is correct.  It \emph{is} possible to manipulate the connector in such a way that it returns to its original state, without moving the object, but only after the object has made \emph{two} full rotations.  If the appropriate manipulation is performed every two rotations, the object can continue to spin indefinitely.

Various physical demonstrations of this striking phenomenon have been devised. In the \emph{belt trick}, the object is the buckle of a belt.  The other end of the belt is held fixed, with the belt initially flat.  The buckle is given two full twists; then the belt can be returned to its original state solely by translating (not rotating) the buckle.  Alternatively, the connector can be a human body.  In the \emph{plate trick}, or \emph{candle dance}, a person holds a plate or a lighted candle in one hand, with a fixed grip, and spins it continuously about a vertical axis, while keeping their feet fixed on the ground.  The arm and body return to their original state every \emph{two} full rotations of the object.
With practice and athleticism, the necessary contortions can be confined primarily to the arm.  Moreover, the trick can be performed with very little horizontal translation movement of the object.  Avoiding vertical translation as well appears more challenging.  The reader is encouraged to explore the many online video demonstrations (easily found from the above search terms), and to try the various tricks for themselves.

One might think of the human arm as a flexible connector akin to a belt, but its motion is more constrained than that.  To a reasonable approximation, it is a linkage composed of a sequence of joints.  Since there are also flexible structures such as blood vessels running along its length, each joint must allow motion only within a certain range.  How many joints are needed, and what kind?  This article is in part motivated by such questions.

The candle dance features in traditional Indonesian and Philippine dances.  The belt trick and a variant involving scissors tethered by multiple strings appear to have been popularized by Paul Dirac, motivated by deep connections to the concept of spin in particle physics.  See e.g.\ \cite{newman} for more information.

Underlying the phenomenon is the celebrated two-to-one map (or double cover) from the $3$-sphere of unit $4$-vectors of reals to the space $SO(3)$ of $3$-dimensional rotations fixing the origin.  The $4$-sphere is simply connected, but the rotation space is not.  One full turn by $2\pi$ of an object about a fixed axis corresponds to a non-contractible loop in the rotation space, which maps back to a path between antipodal points on the sphere.  A double turn by $4\pi$ maps to a great circle of the sphere, and is therefore contractible.  The belt or connector may be viewed as a path in the rotation space, encoded by the orientations of elements along its length.  The untwisting operation sweeps out a surface in the rotation space which corresponds to a contraction.  The rotation space has a group structure as well as a topological structure.  If $3$-sphere is parameterised by unit quaternions, then the map becomes a continuous group homomorphism.  In this setting, elements of the sphere are known as \emph{spinors}.  Detailed accounts of the mathematical background may be found for example in \cite{hanson,staley,penrose,waving}.

%

The purpose of this article is to introduce a new physical demonstration of the phenomenon discussed above: a mechanical device that performs the plate trick.  We call it the spinor linkage.  Shown in \cref{trunk}, the device takes the form of a mechanical arm that plays a role analogous to the human arm.  It consists of 13 rigid bodies (each shown in one colour), connected in sequence by 12 hinge joints.  Each hinge joint permits its two incident bodies to pivot about a fixed axis relative to themselves, through an interval of angles of length $90^\circ$.  (In a natural frame of reference suggested by the shapes of the rigid bodies, each interval is $[-45^\circ,45^\circ]$).  The body at one end (near the top of the figure) is held fixed, while the body at the other end (the small cube near the center of the figure) is able to spin continuously about a fixed (vertical) axis, while keeping the same location.  Moreover (and this is the key innovation), this motion is produced by a mechanism inside the arm, operated simply by turning a handle.  In fact, the control mechanism turns out to be very simple, and requires only one further moving part.  The entire device has exactly one mechanical degree of freedom, and its motion forms a loop in the configuration space.

\begin{figure}
  \includegraphics[width=.65\linewidth]{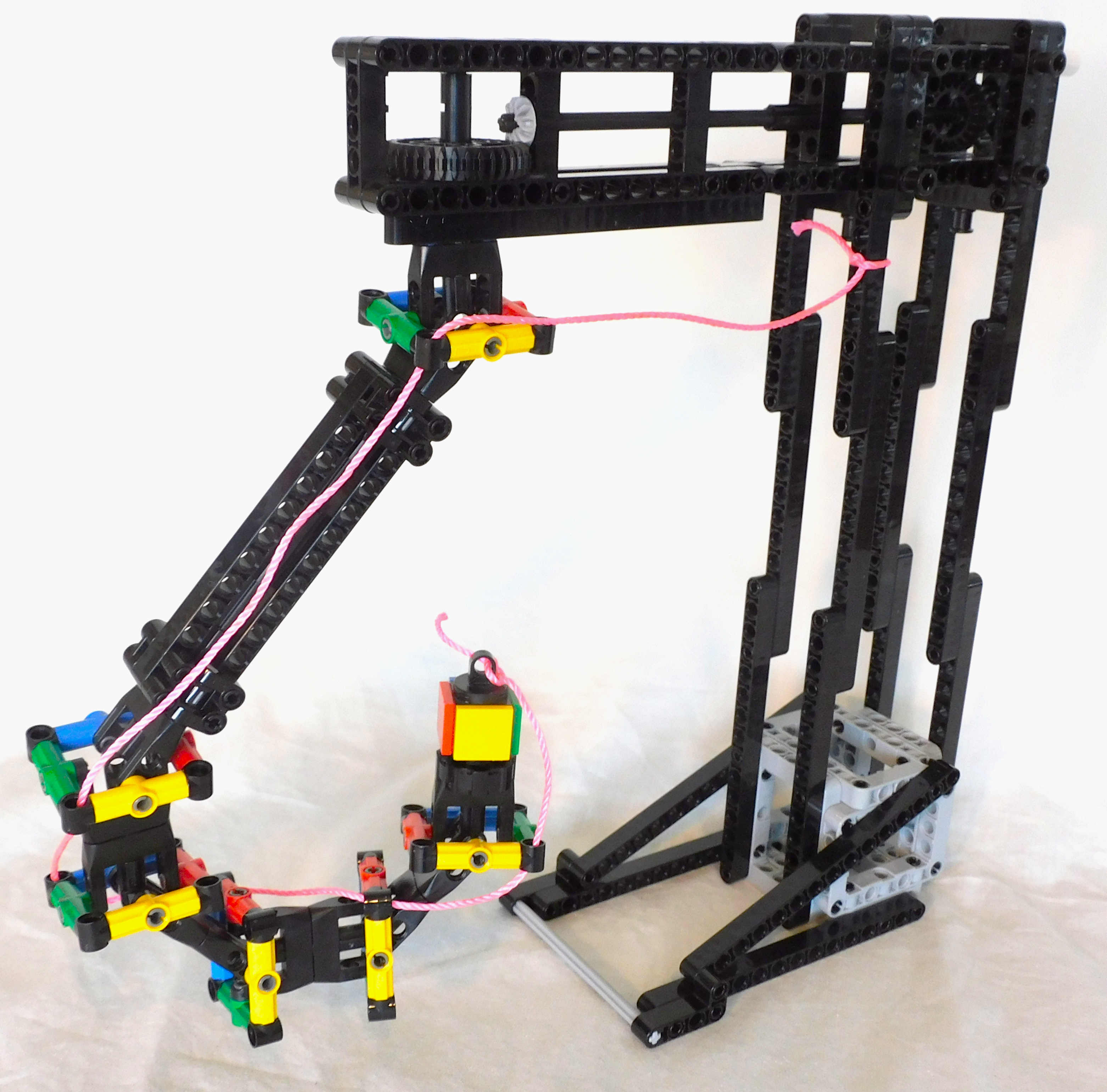}
  \caption{Implementation in LEGO, with string added.}
  \label{lego}
\end{figure}

\cref{lego} shows a working physical model of the spinor linkage, made by the author from LEGO pieces.  See \cite{vid} for a video.  As seen here, it is possible to run a piece of string along the outside of the arm in such a way that it flexes along with the arm.  One end of the string can be tied off to the fixed external frame, while the other end is tied to the cube and rotates continuously, without the string ever getting twisted up.  The version depicted here has a different colour scheme which helps to illustrate some features of the motion.  The arm has a roughly square cross-section, and each of its four sides has its own colour.

Even when one sees the plate trick or its variants in action and understands the mathematics, the phenomenon remains quite mysterious.  It is far from easy to understand intuitively how the motion is possible, and what is really going on.  As we will explain, the new mechanism offers some perspectives which may be helpful.

The spinor linkage was discovered by the author via experimentation with LEGO pieces, after hearing about the plate trick and its mathematical underpinnings in a beautiful lecture series by Richard Schwartz given at the Illustrating Mathematics Programme at ICERM in 2019.  LEGO is an extremely useful platform for exploring mechanisms, and it turns out to be ideally suited to the particular problem at hand.  In an appendix we provide full instructions for the reader to make their own LEGO implementation.

\begin{figure}[h]
  \includegraphics[width=.35\linewidth]{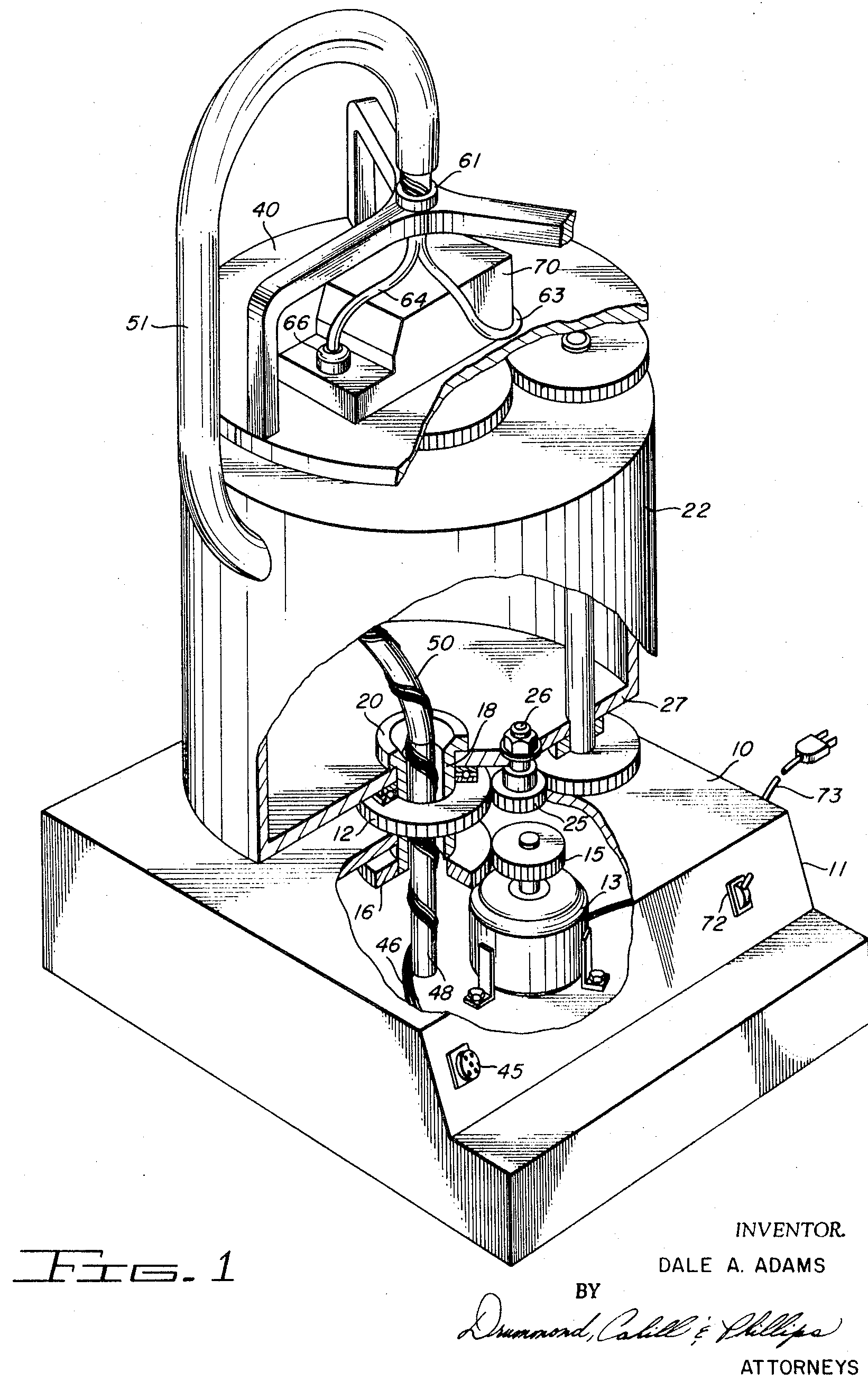}
  \caption{Adams' anti-twist mechanism; patent US 3586413.}
  \label{twist}
\end{figure}

The closest relative of the spinor linkage that we are aware of is the anti-twist mechanism patented in 1971 by Dale A.\ Adams \cite{adams} -- see \cref{twist}.  This uses gearing to guide a flexible conduit in such a way that its free end spins without twisting.  Note the similarity in shape to \cref{trunk}.  We briefly discuss anti-twist mechanisms later.

\section{The mechanism}

\begin{figure}
\begin{minipage}{0.5\textwidth}
\centering
\includegraphics[width=.9\linewidth]{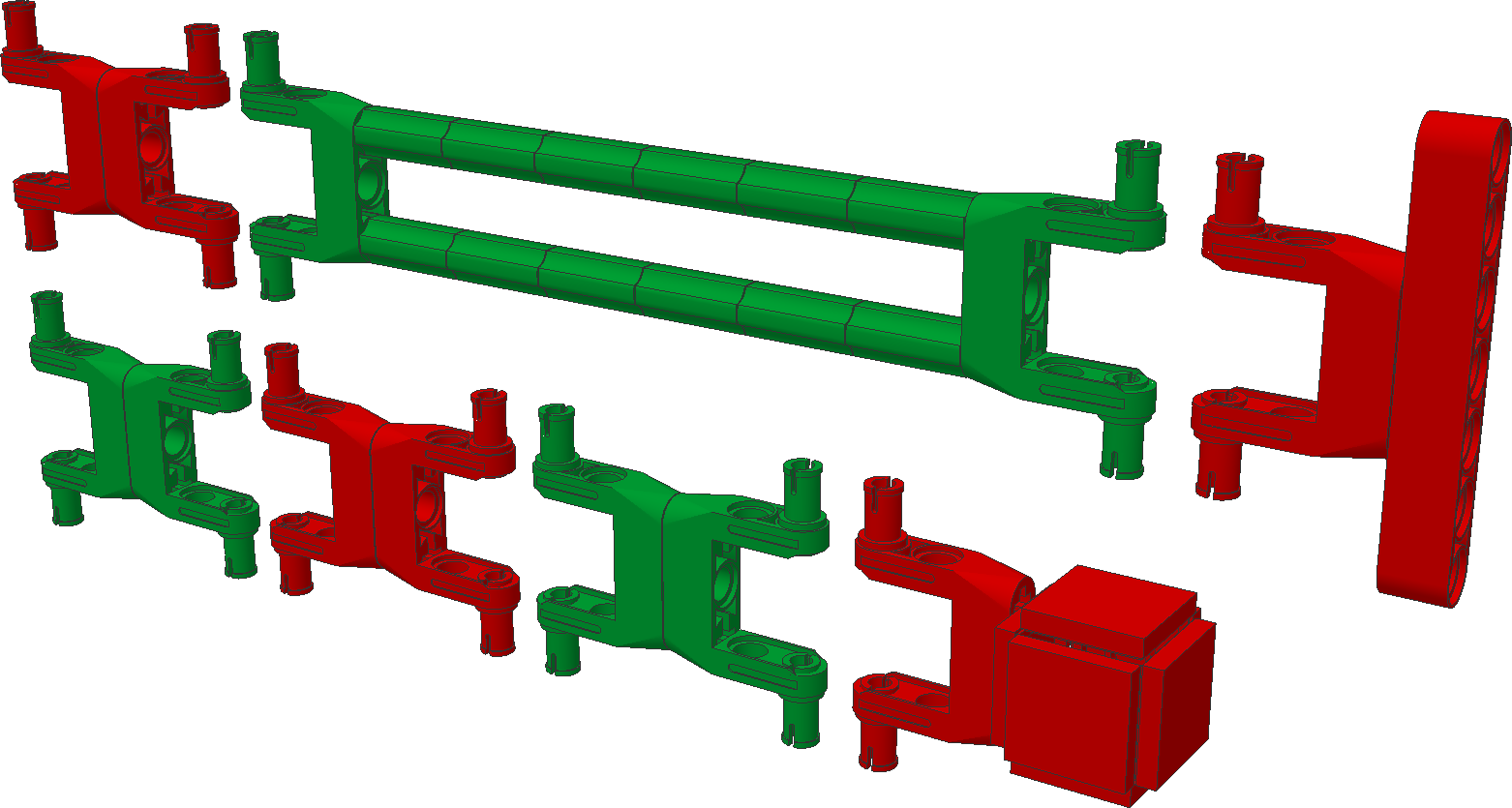}
\subcaption{Flat units}
\label{flat}
\vspace{5mm}
\includegraphics[width=.7\linewidth]{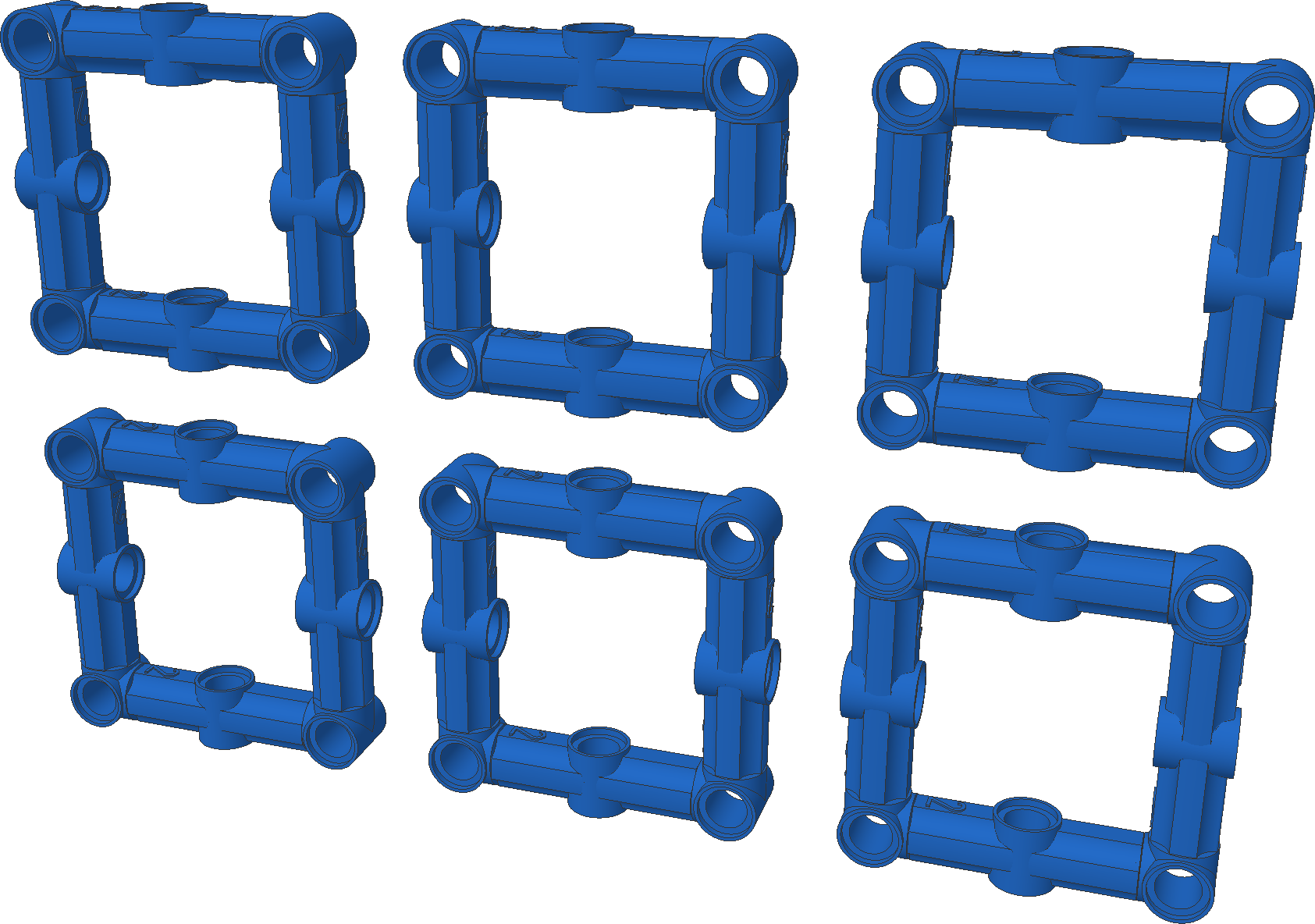}
\subcaption{Rings}
\label{rings}
\end{minipage}\hfill
\begin{minipage}{0.25\textwidth}
\centering
\includegraphics[width=.9\linewidth]{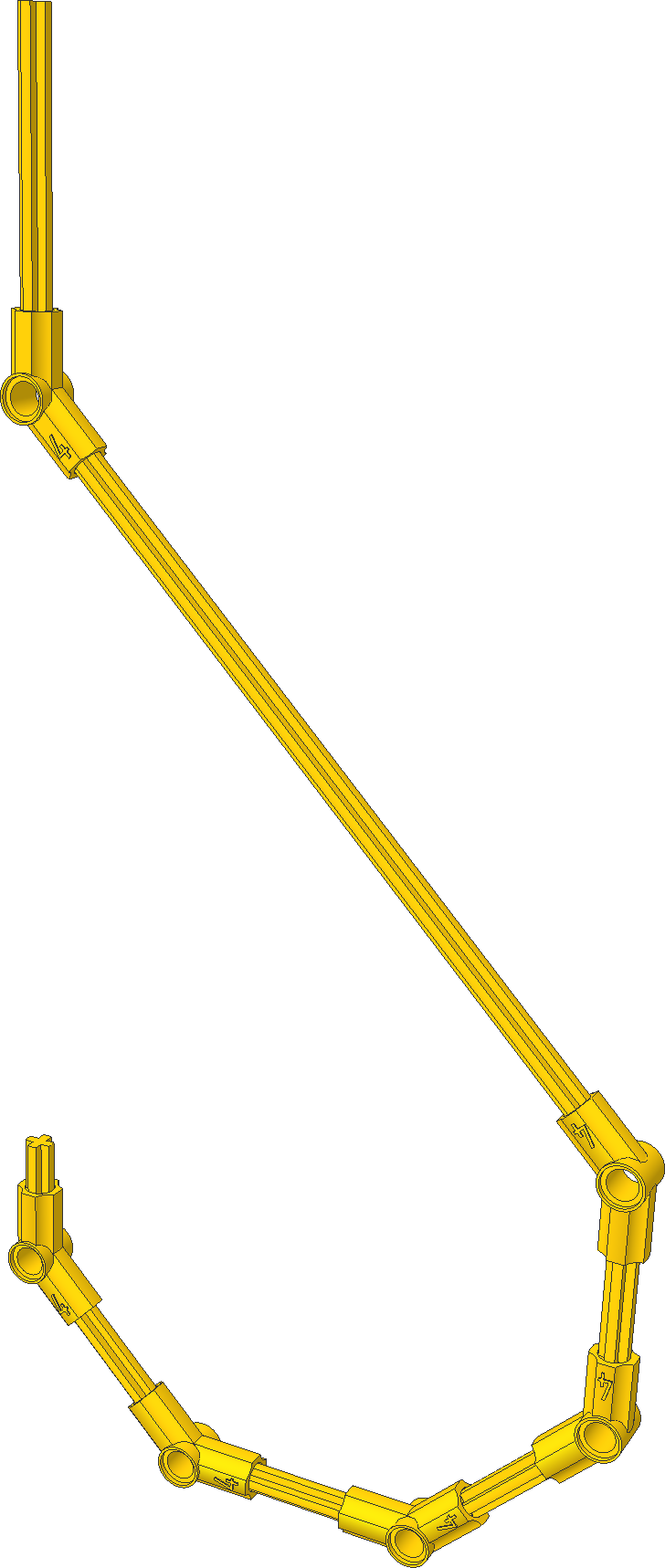}
\subcaption{Core}
\label{hook}
\end{minipage}
\caption{The rigid components.}
\label{comps}
\end{figure}

As mentioned above, the spinor linkage comprises 14 rigid bodies in total.  The 7 ``flat'' units in \cref{flat} alternate with the 6 identical rings in \cref{rings} to form the outer arm.  These 13 bodies are connected to each other in sequence via hinge joints.  The final body is a single rigid hook-shaped shaft (\cref{hook}) which runs through the inside of the arm through a series of holes in the flat units, and provides the control mechanism.  This shaft is free to rotate in the holes, but cannot slide lengthways.
\begin{figure}
\centering
  \includegraphics[align=t,width=.28\linewidth]{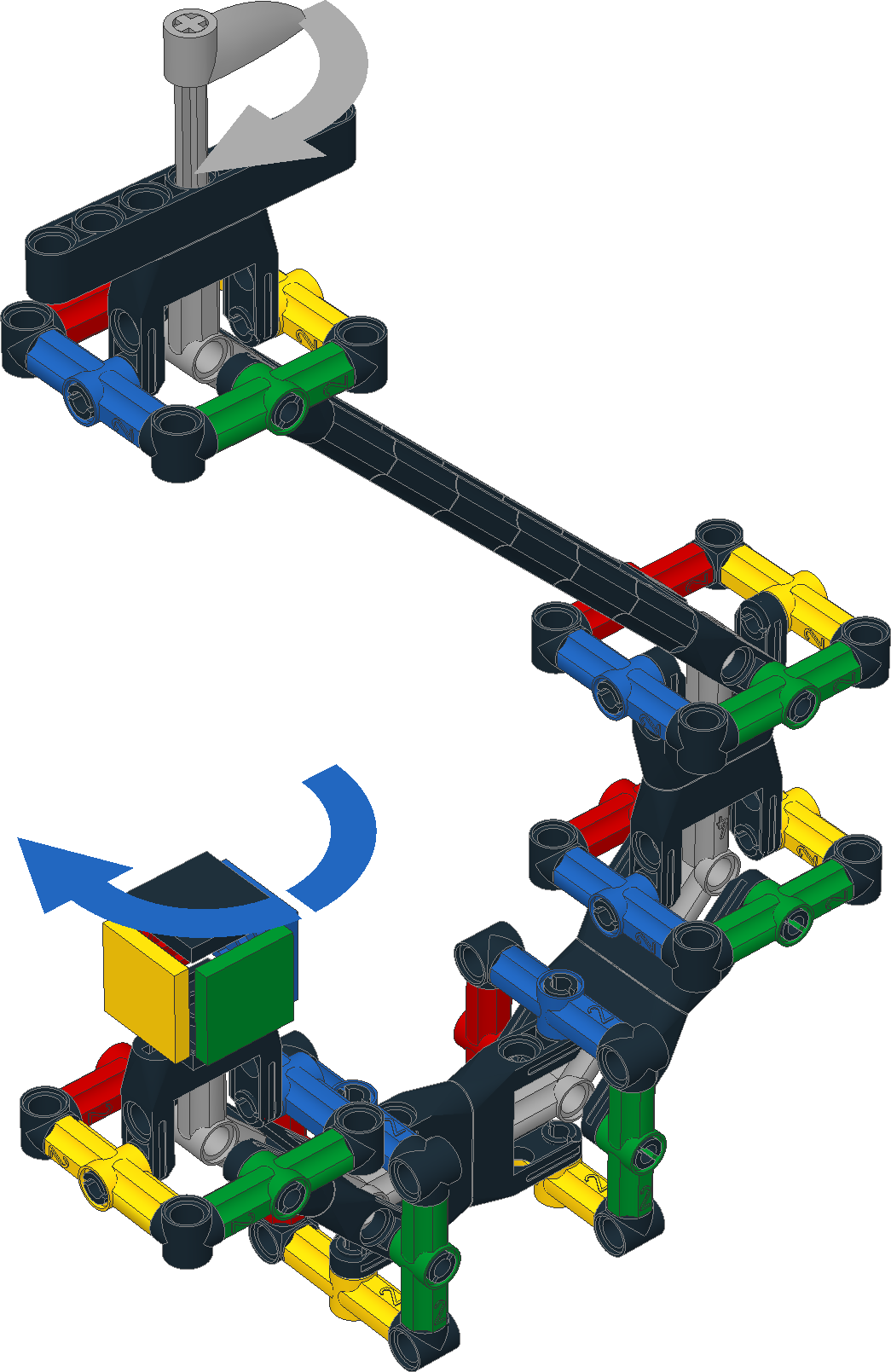}
  \includegraphics[align=t,width=.27\linewidth]{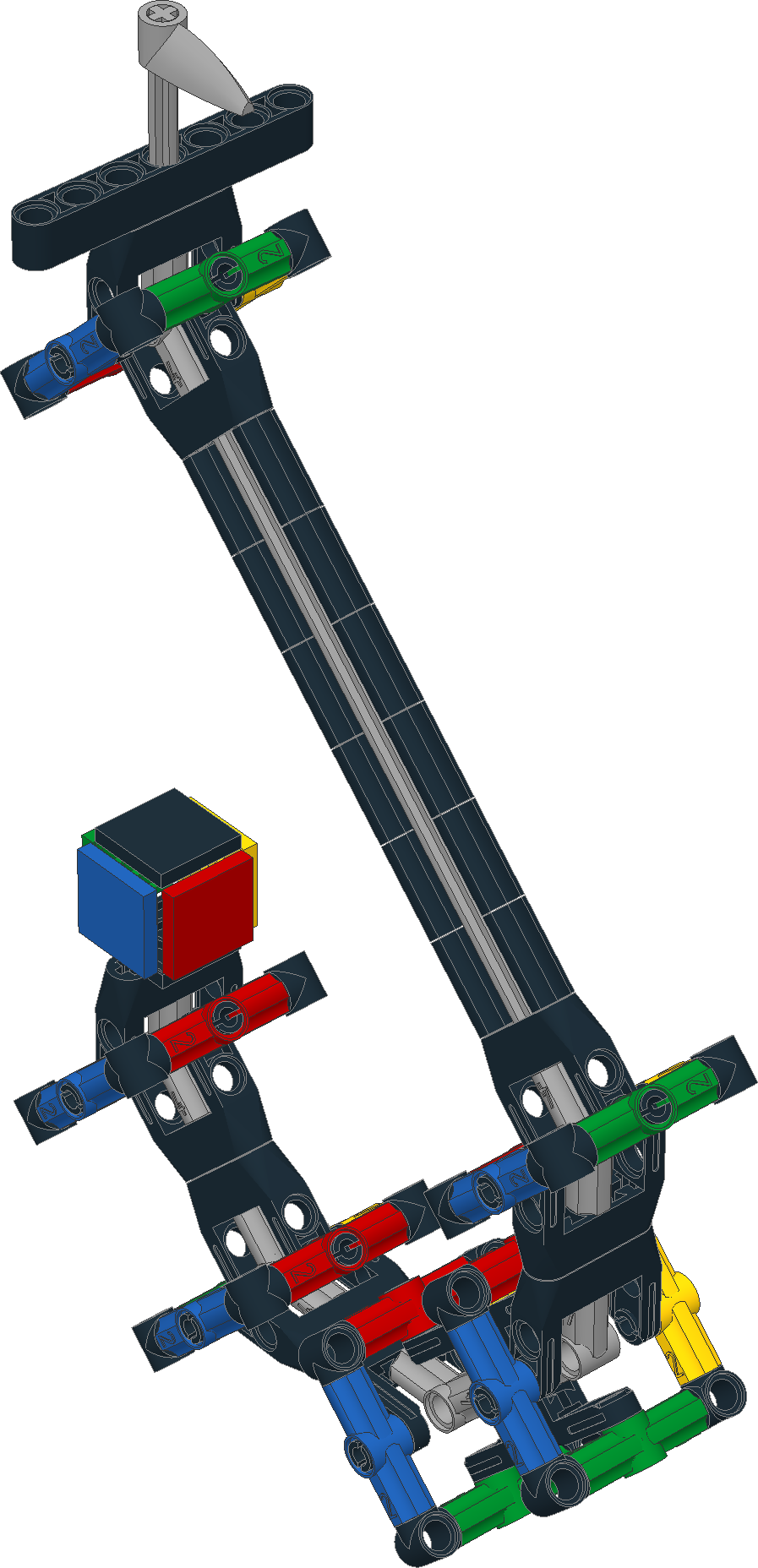}
  \includegraphics[align=t,width=.32\linewidth]{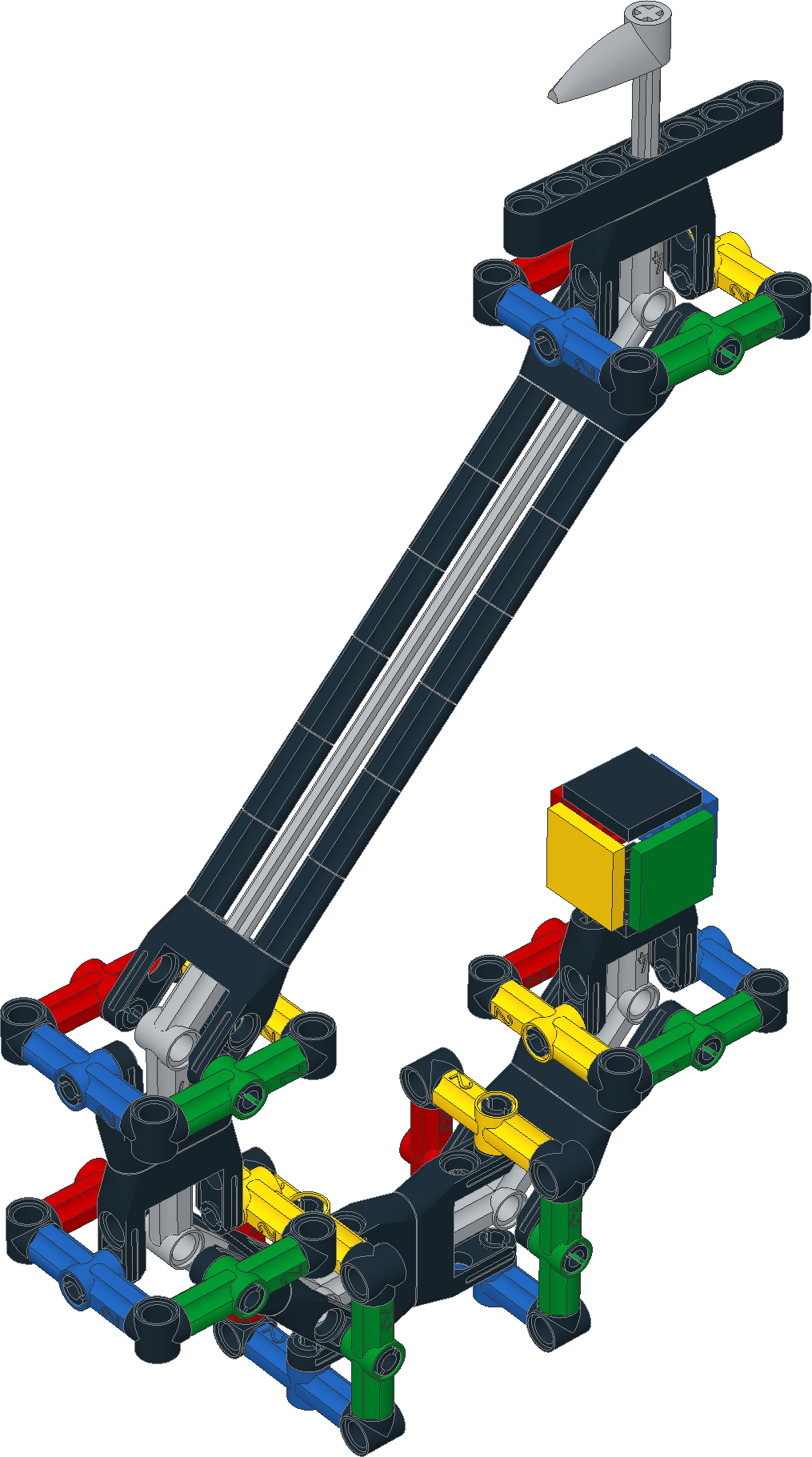}
  \caption{Three snapshots of the motion: the control shaft performs two quarter turns clockwise causing the cube to perform two half turns clockwise.}
  \label{snaps}
\end{figure}

In operation, the uppermost flat unit is held rigidly in place (by a external frame), and the hook-shaped control shaft is rotated around a vertical axis, by means of its uppermost protruding end.  Naturally this results in the arm swinging around together with the hook.  The other end of the hook lies on its rotation axis, inside the cube which forms the other end of the arm.  Hence the cube remains at the same location.  However, the arm changes shape locally.  Its parts roll around the hook along its length, which results in the cube also rotating in place around a vertical axis.

It is crucial that the lower end of the hook is not connected rigidly to the cube.  Instead it is free to rotate in a vertical hole in the cube.  Indeed, the hook and cube rotate at different speeds.  The cube rotates at exactly twice the speed of the hook, and in the same direction.  This is intimately related to the underlying mathematics of the plate trick as discussed earlier.  When the cube has made one full rotation, the arm is not in its original state (indeed, that would be impossible).  Instead it has the opposite orientation.  Two full rotations of the cube restore the arm to its original position.

Three snapshots of the motion are shown in \cref{snaps}.  Here the four sides of the arm are again assigned different colours.  The control shaft is rotated through $180^\circ$ (in two steps of $90^\circ$), causing the cube to rotate by $360^\circ$ (in two steps of $180^\circ$).  Watching the video \cite{vid} is highly recommended.

\section{How it works}

\begin{figure}
\begin{subfigure}{.25\textwidth}
  \centering
  \includegraphics[height=.45\linewidth]{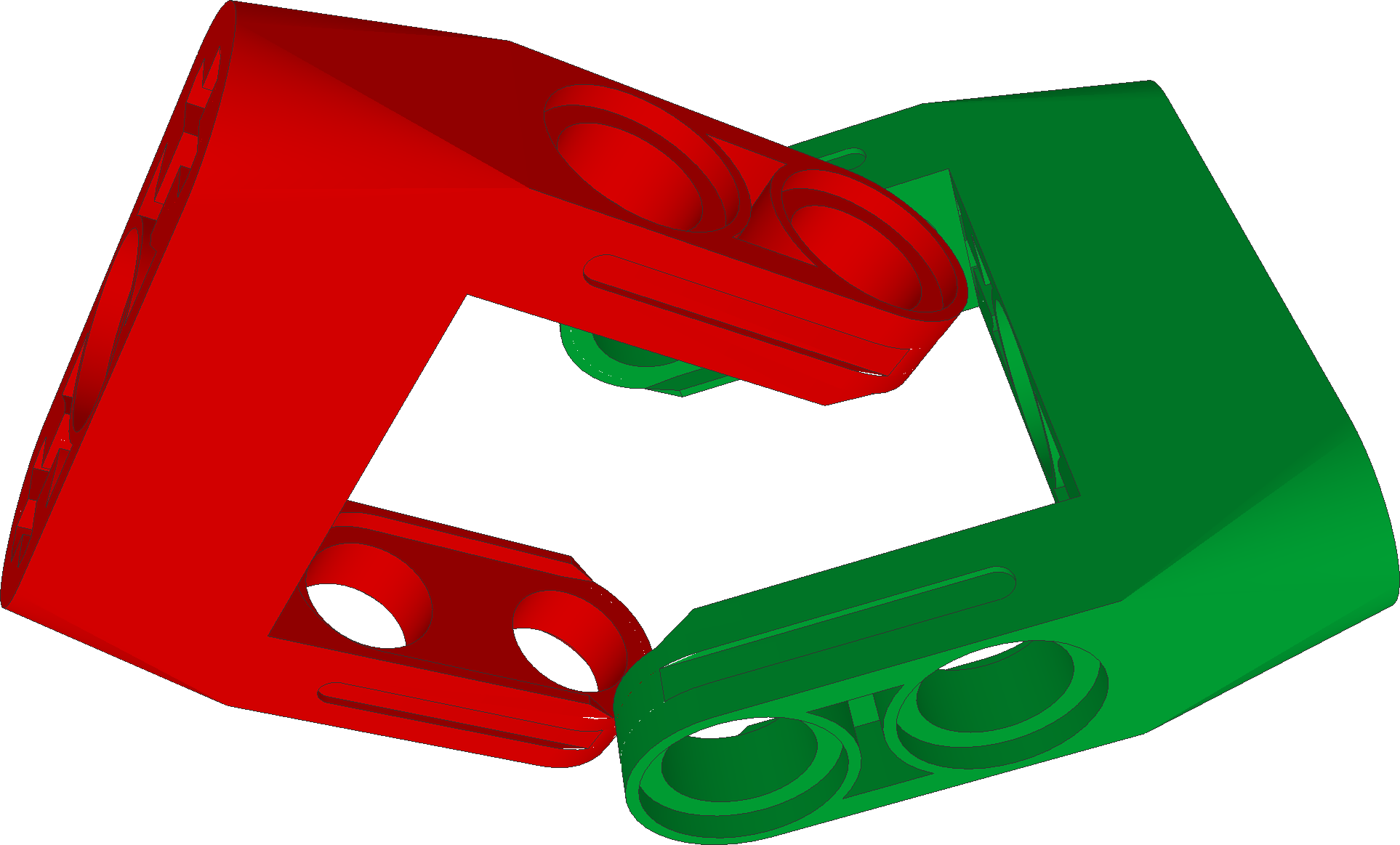}
  \caption{Yokes}\label{yokes1}
\end{subfigure}
\hfill
\begin{subfigure}{.25\textwidth}
  \centering
  \includegraphics[height=.45\linewidth]{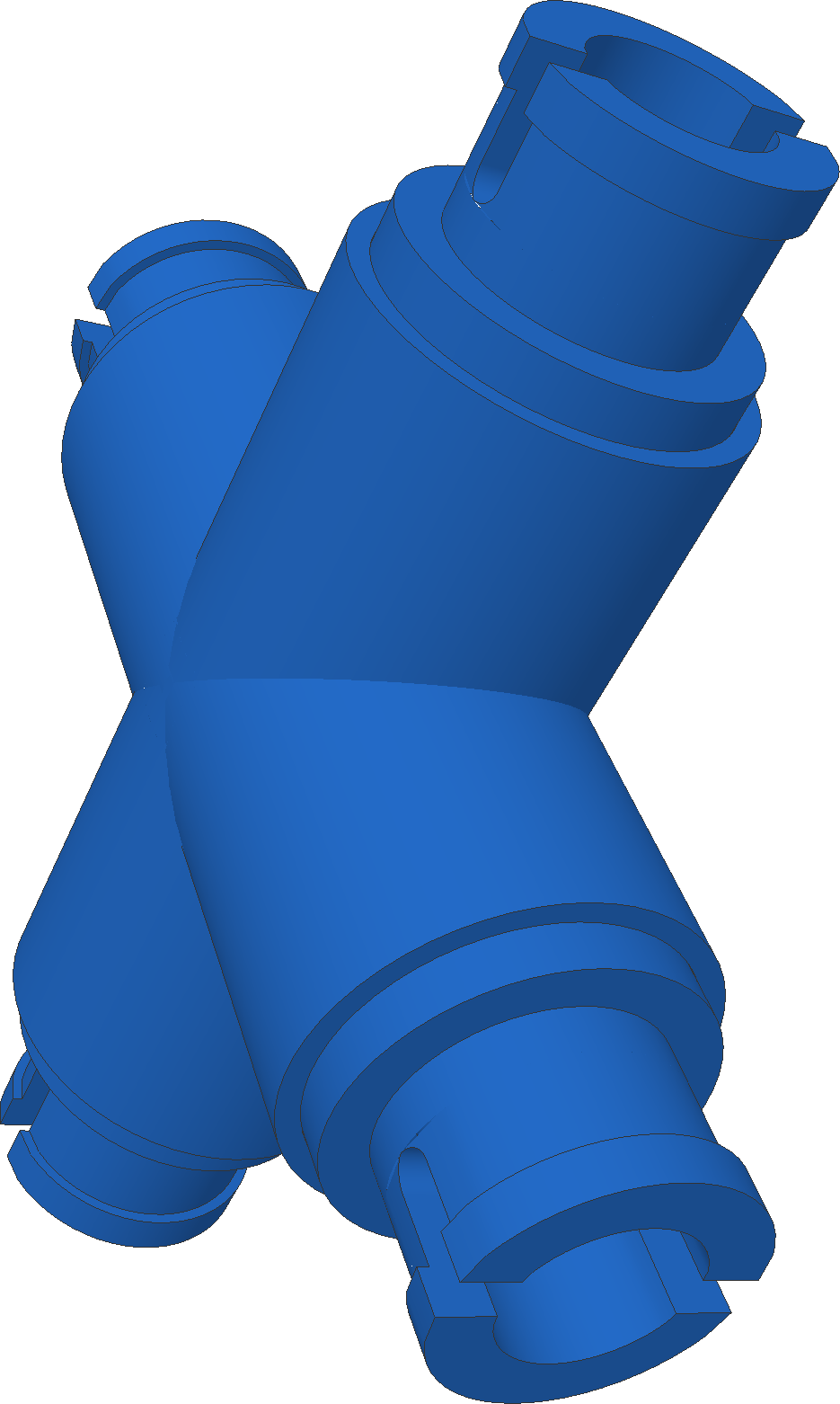}
  \caption{Cross}\label{cross}
\end{subfigure}
\hfill
\begin{subfigure}{.25\textwidth}
  \centering
  \includegraphics[height=.47\linewidth]{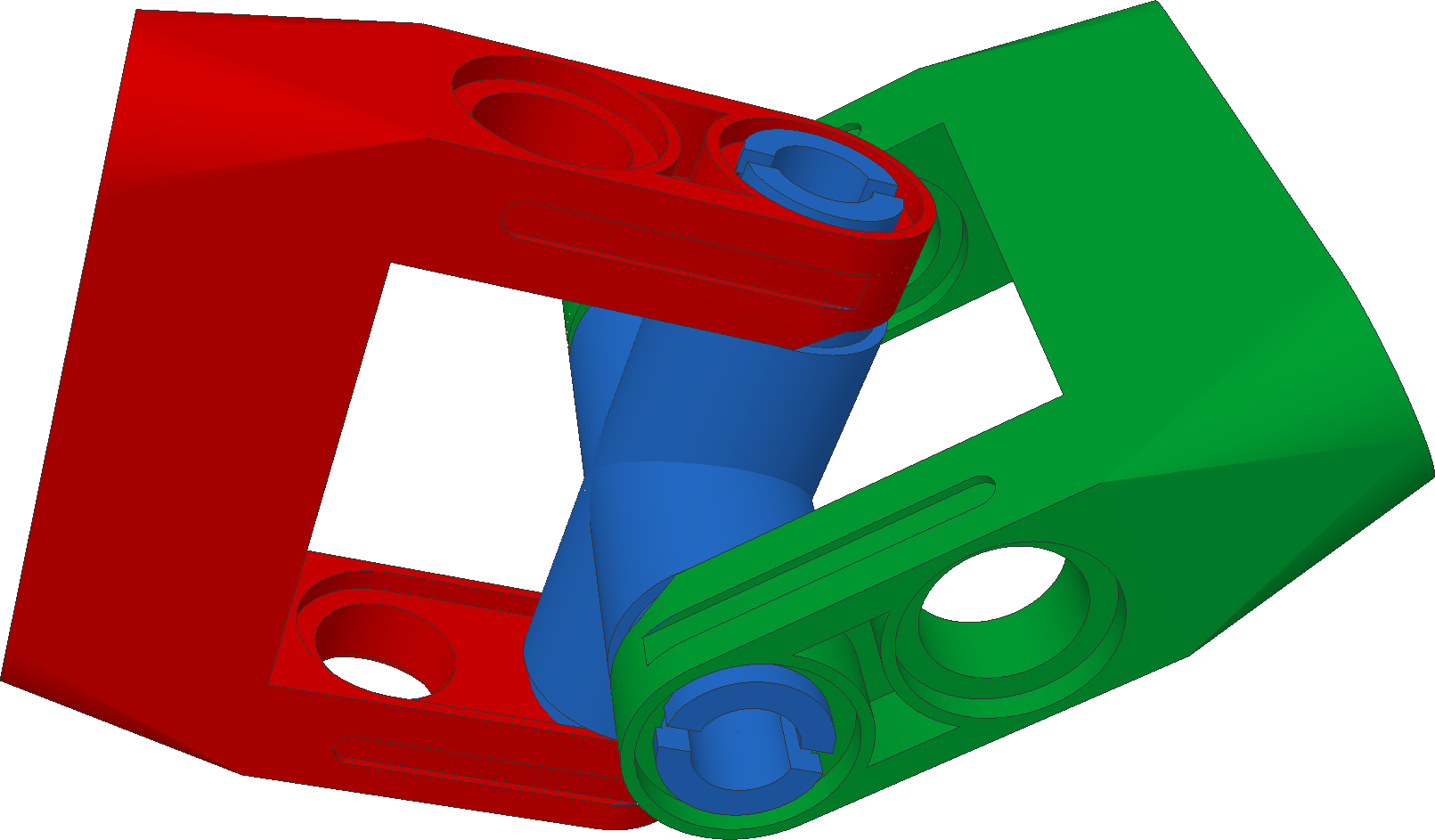}
  \caption{Complete joint}
\end{subfigure}
\newline
\begin{subfigure}{.5\textwidth}
  \centering
  \includegraphics[width=\linewidth]{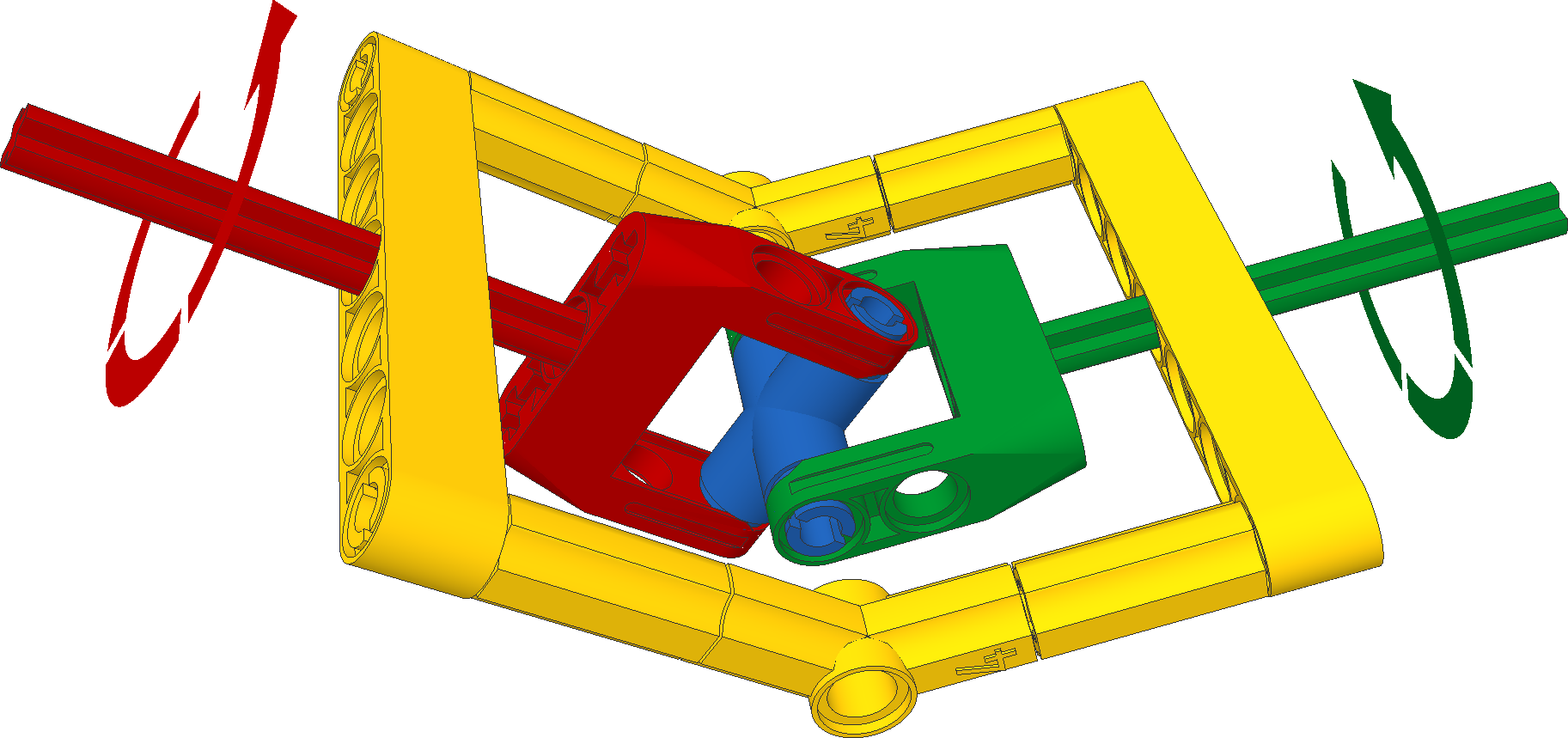}
  \caption{Transmission of rotary motion between shafts, constrained to a fixed angle by a frame}
  \label{frame}
\end{subfigure}
\caption{A conventional universal joint.}
\label{conventional}
\end{figure}

Our starting point is the \emph{universal joint}, as shown\footnote{We employ diagrams constructed using CAD tools designed for LEGO parts to illustrate the workings of the various mechanisms.  However, these explanatory diagrams do not in general correspond to fully functional physical LEGO constructions.  Some involve non-existent parts, or unrealistic connections between parts.  On the other hand, the instructions in the appendix are for a complete fully working LEGO model.} in \cref{conventional}.  This well-known mechanism consists of three rigid bodies: two \emph{yokes} (\cref{yokes1}), each of them connected to a central \emph{cross} (\cref{cross}) via a pivot (or hinge joint).  The axes of the two pivots intersect each other at right angles at the center of the cross.  In typical applications, the two yokes are extended to form shafts, with their center lines each also passing through the center point of the cross.  The universal joint then enables rotary motion to be transmitted from one shaft to the other, while the angle between the shafts can be varied.  In \cref{frame}, the shafts are shown running through holes in an external frame, so that they are free to rotate, but the angle between them is fixed.

The universal joint was analysed by Robert Hooke in the 17th century.  We note in passing that it is not a constant-velocity joint.  That is to say, if the angle between the shafts is non-trivial, then rotating one shaft at a uniform speed results in a non-uniform rotation of the other.  However, if two universal joints are joined by an intermediate shaft then the non-uniformity is cancelled out, provided the two joints form equal angles, and their relative phases are chosen appropriately.  This configuration is known as a double Cardan joint.

\begin{figure}
\centering
\begin{subfigure}{.24\textwidth}
  \centering
  \includegraphics[height=.6\linewidth]{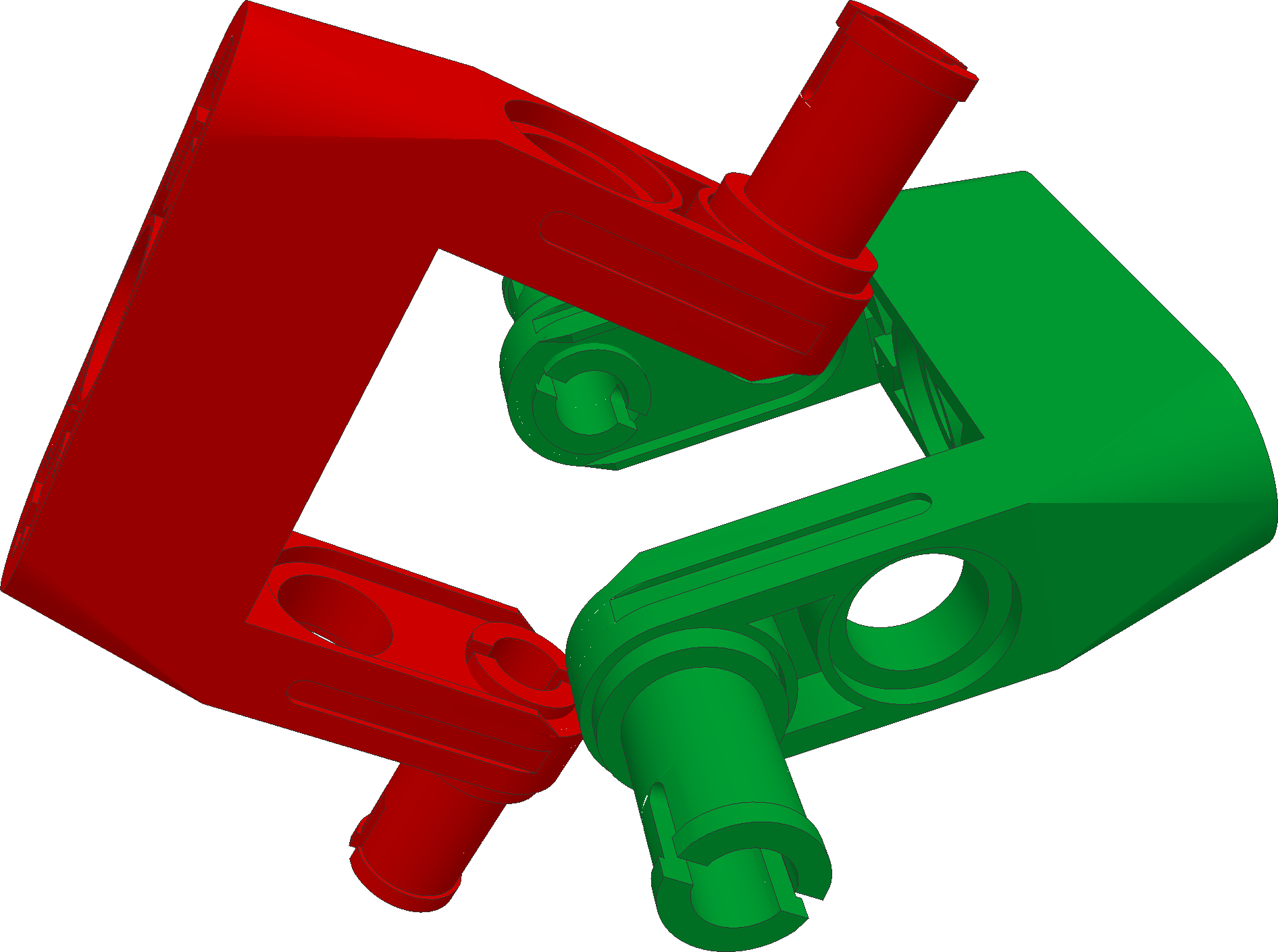}
  \caption{Yokes}
  \label{yokes2}
\end{subfigure}
\hfill
\begin{subfigure}{.24\textwidth}
  \centering
  \includegraphics[height=.7\linewidth]{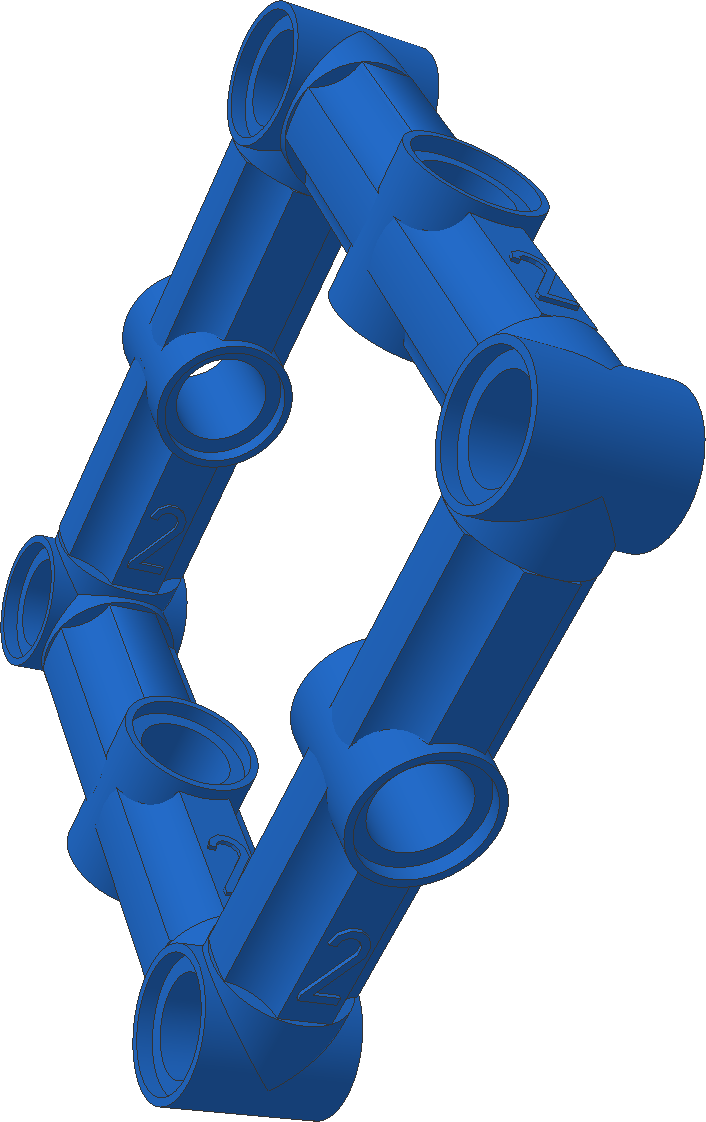}
  \caption{Ring}
  \label{ring}
\end{subfigure}
\hfill
\begin{subfigure}{.24\textwidth}
  \centering
  \includegraphics[height=.75\linewidth]{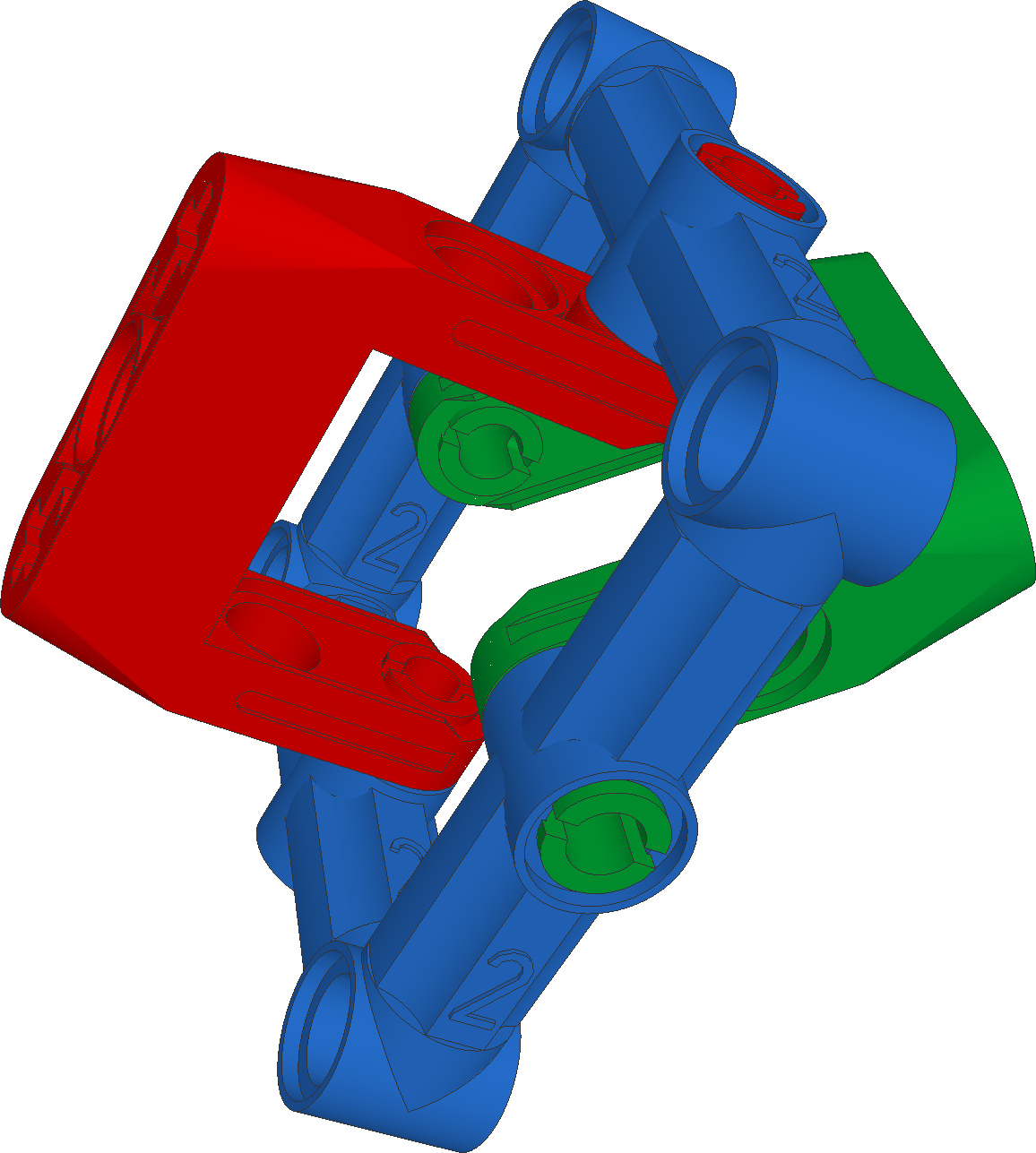}
  \caption{Complete joint}
\end{subfigure}
\begin{subfigure}{.25\textwidth}
  \centering
  \includegraphics[width=\linewidth]{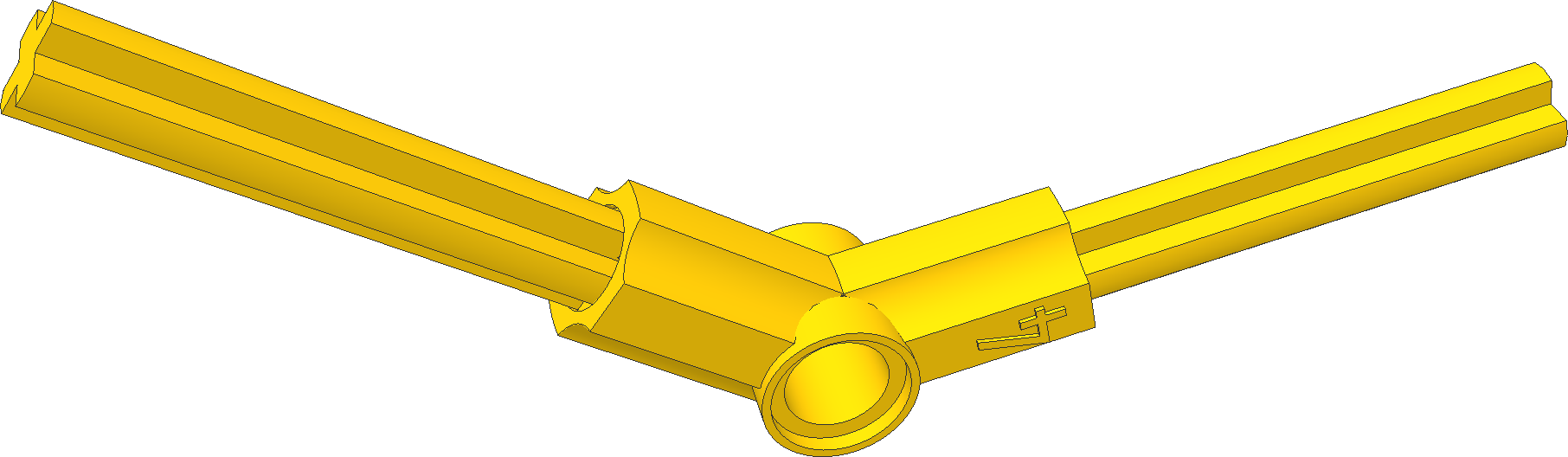}
  \caption{Central core}
  \label{core}
\end{subfigure}

\begin{subfigure}{.4\textwidth}
  \centering
  \includegraphics[width=\linewidth]{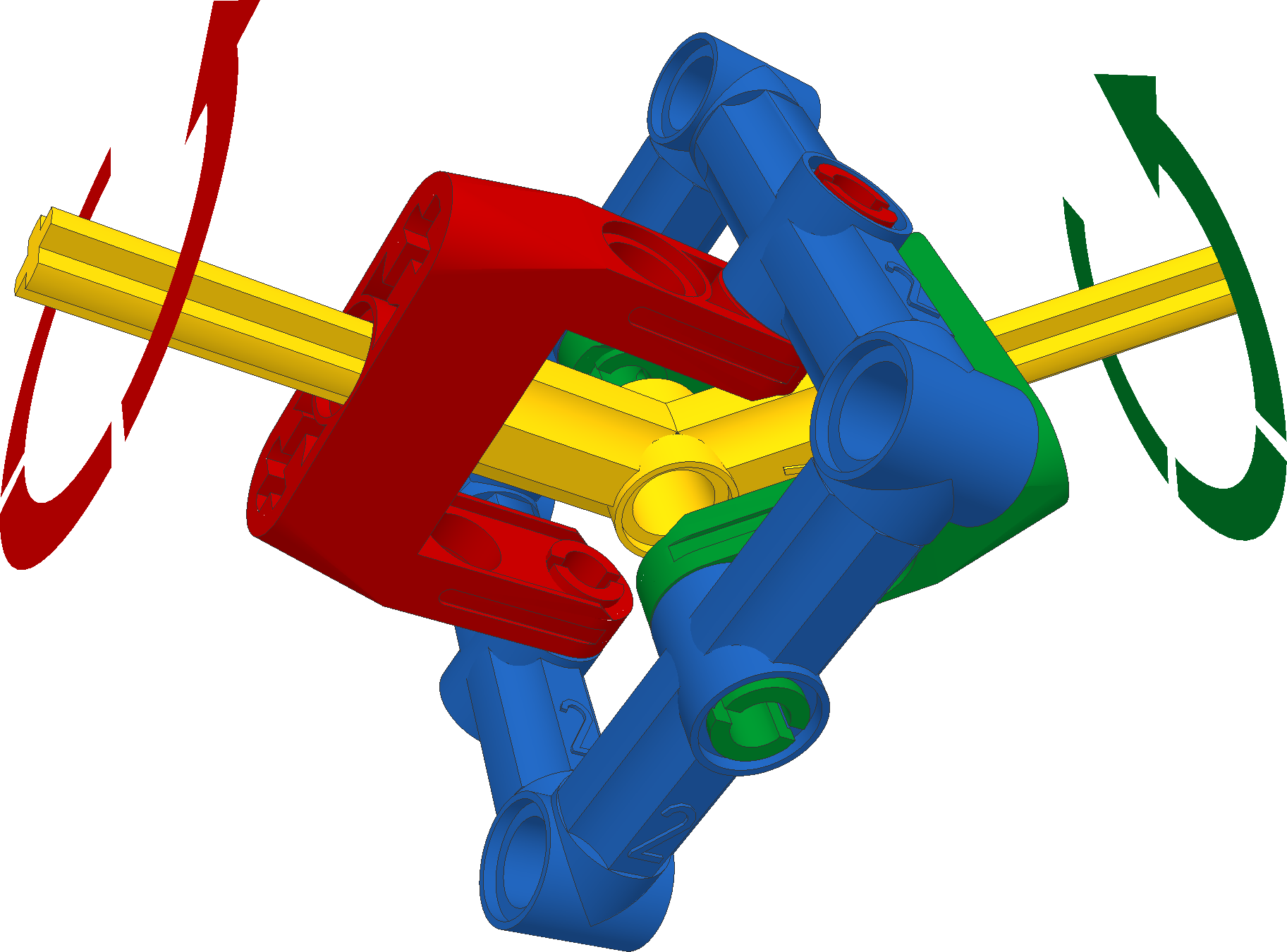}
  \caption{Rotation around the core}
  \label{rotate}
\end{subfigure}
\caption{The inside-out universal joint.}
\label{joint}
\end{figure}

The key ingredient of our mechanism is an inside-out version of a universal joint, shown in \cref{joint}.  Here the cross is replaced by a ring (\cref{ring}) located \emph{outside} the two yokes, rather than inside, but still connected to them via pivots with the same axes as before.  This does not change the essential mechanics, but permits new possibilities for adjacent structures by freeing up space at the center.  In particular, we introduce a central rigid core as shown in \cref{core}, consisting of a shaft with a bend at the center, which is situated so that the two ends pass through holes in the two yokes.  This allows the yokes to rotate around the two shafts (\cref{rotate}), with rotation transmitted from one yoke to the other as before.  The core plays precisely the same role as the frame in \cref{frame}.

In our implementation the shafts fit snugly in the holes, and the core is equipped with a collar at its center that is wider than the shafts.  The collar cannot fit through the holes in the yokes, but abuts them.  Consequently the shafts cannot slide in the axial direction through the holes, and the motion of the mechanism has exactly one degree of freedom.

\begin{figure}
  \centering
  \includegraphics[width=.45\linewidth]{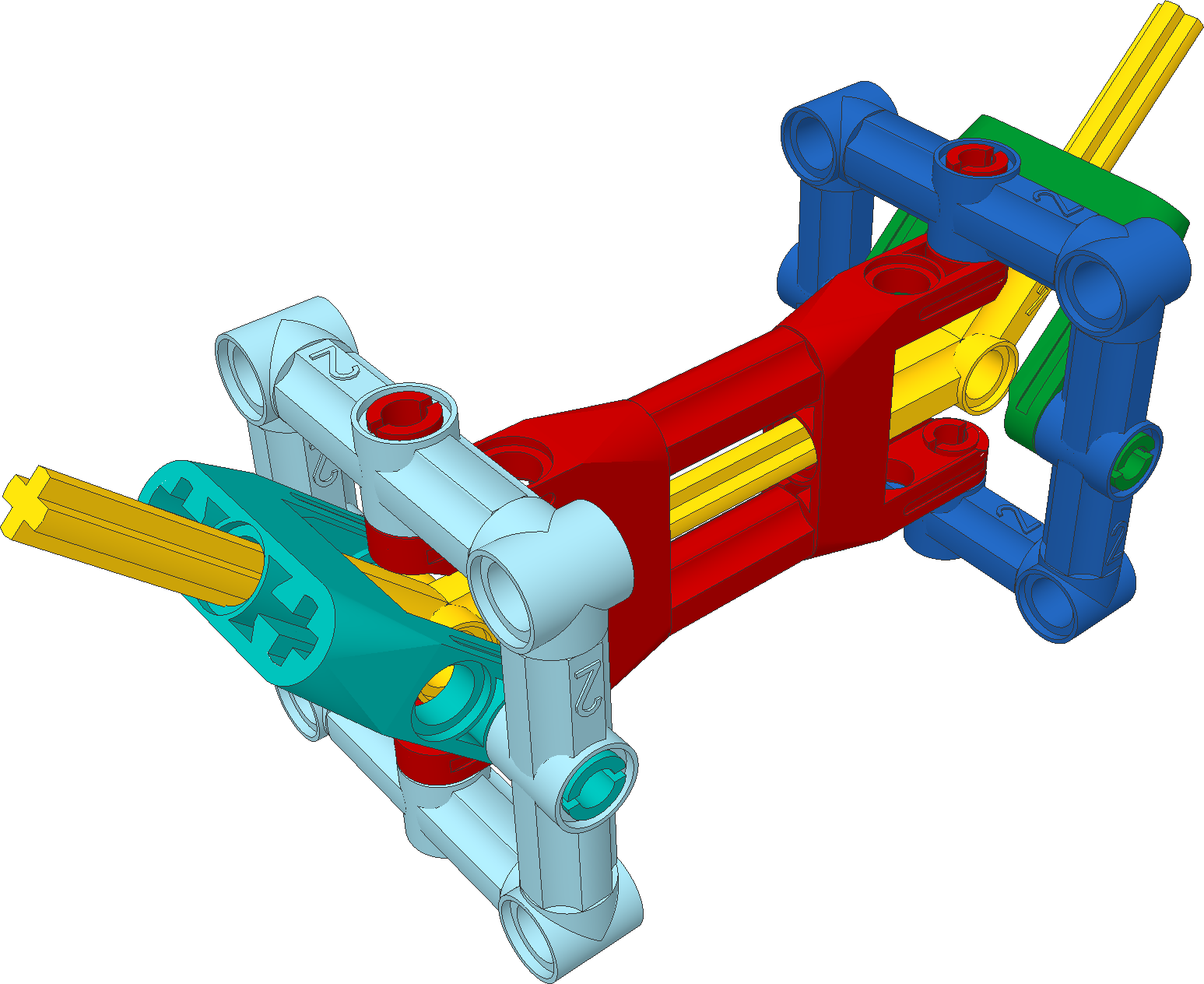}
  \caption{Two inside-out universal joints connected.}
  \label{two}
\end{figure}

Now consider two inside-out universal joints joined together as in \cref{two}.  We fuse together one yoke from each joint to form a single rigid body, and we fuse the two cores to make a single rigid shaft containing two bends (each $45^\circ$ in this case).  The resulting mechanism has 5 rigid components, and its motion has one degree of freedom.  The yokes and rings roll around the core together.  One can view it as a method of transferring rotary motion around two bends, between the two outer yokes.  Since the two bends have equal angles and the two fused yokes are aligned in the manner shown, with their pivot axes parallel, it amounts to a constant-velocity joint.

\begin{figure}
  \centering
  \includegraphics[width=.15\linewidth]{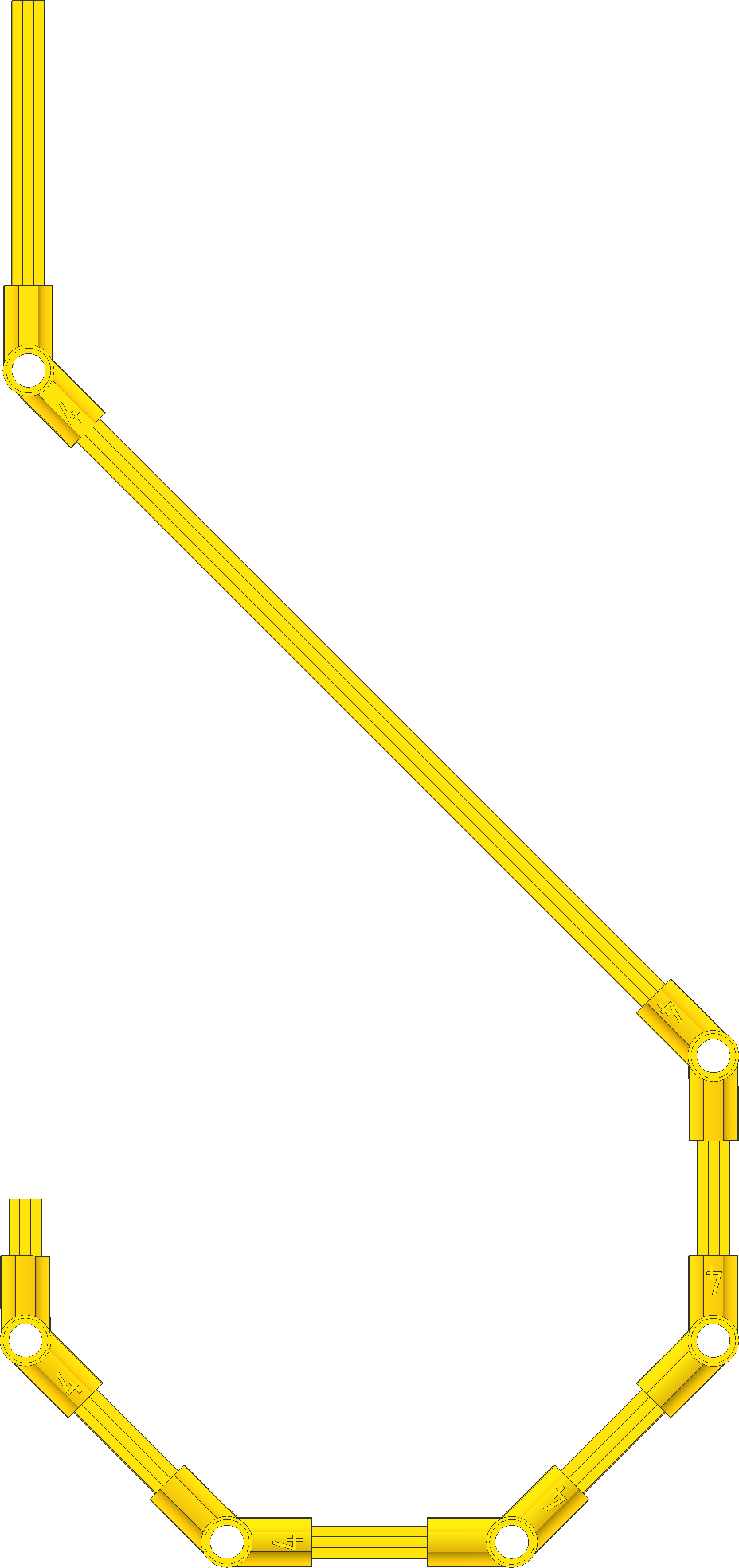}
  \caption{The core.}
  \label{orth}
\end{figure}
Our mechanism now simply consists of six inside-out universal joints connected  sequentially in the same manner as above.  Each bend angle is $45^\circ$, and all the shafts lie in one plane, with directions and lengths chosen so that they form the hook-shaped core shown in \cref{core} -- shown also in \cref{orth} in a planar view without perspective.  The two straight shafts at the ends of the hook are segments of the same line, and point in the same direction.  Since the number of joints is even, and they are oriented in the same manner as in \cref{two}, the entire mechanism functions as a constant-velocity joint.

\section{Understanding the motion}

It is helpful to think of the arm in two sections: the lower part which is roughly semicircular, and the upper part which is primarily straight with two shorter parallel appendages.  The motion of each section individually is relatively easy to understand.

\begin{figure}
  \centering
  \includegraphics[width=.5\linewidth]{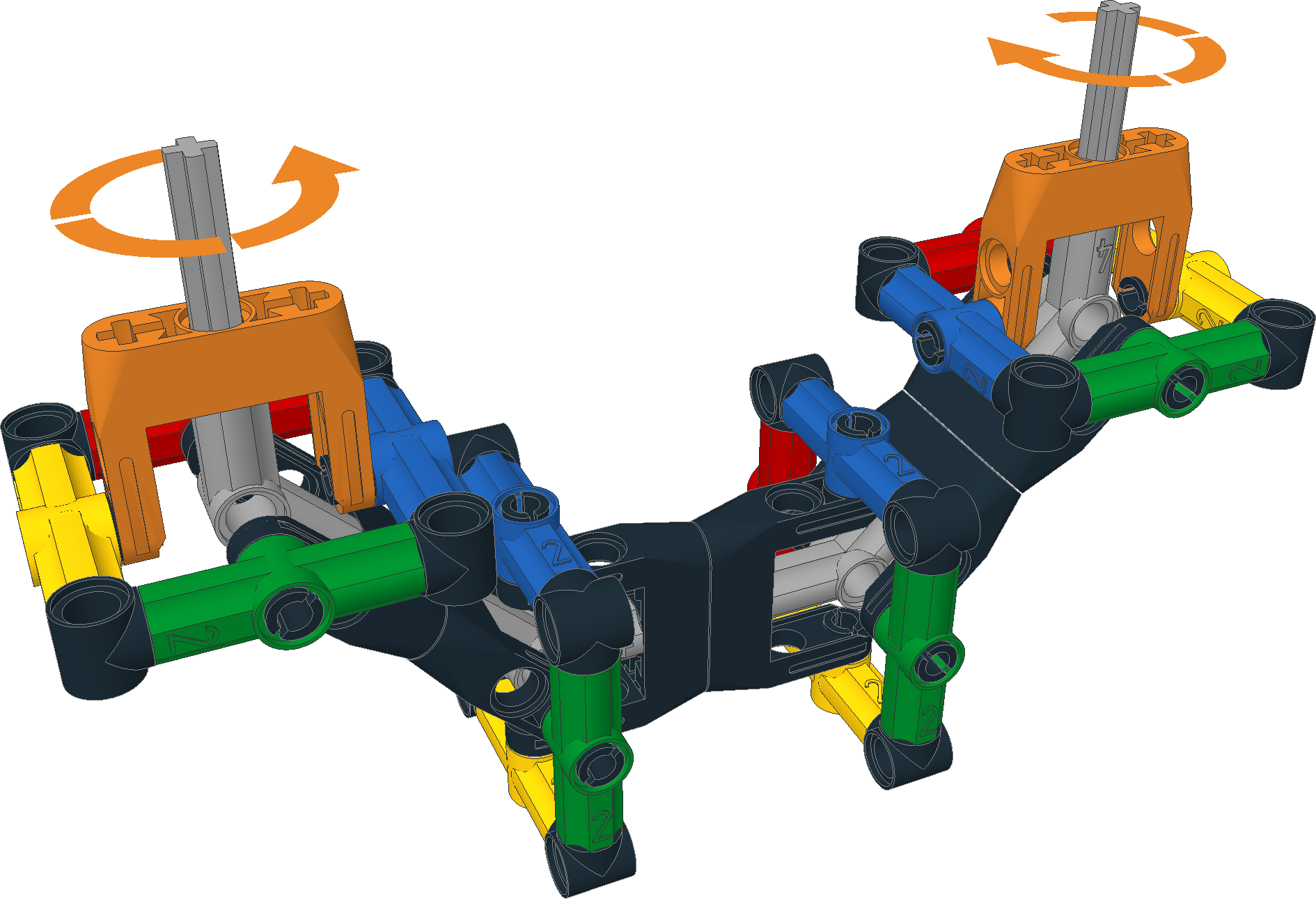}
  \caption{Motion of the lower section.}
  \label{curve}
\end{figure}

\begin{figure}
\centering
  \begin{subfigure}{.45\textwidth}
  \centering
  \includegraphics[width=\linewidth]{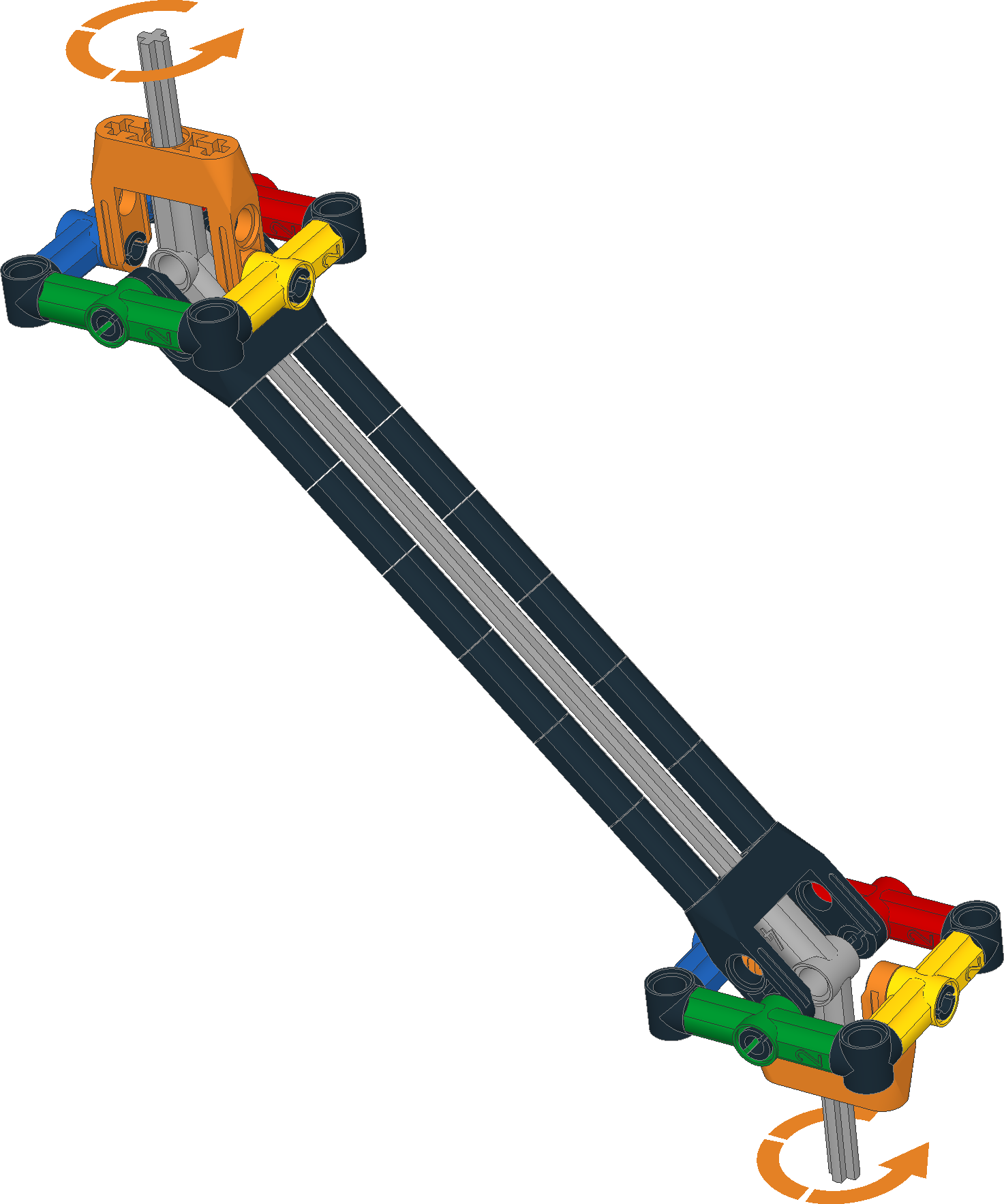}
  \caption{Rotating in place}
  \label{straight2}
  \end{subfigure}
\hfill
\begin{subfigure}{.4\textwidth}
  \centering
  \includegraphics[width=\linewidth]{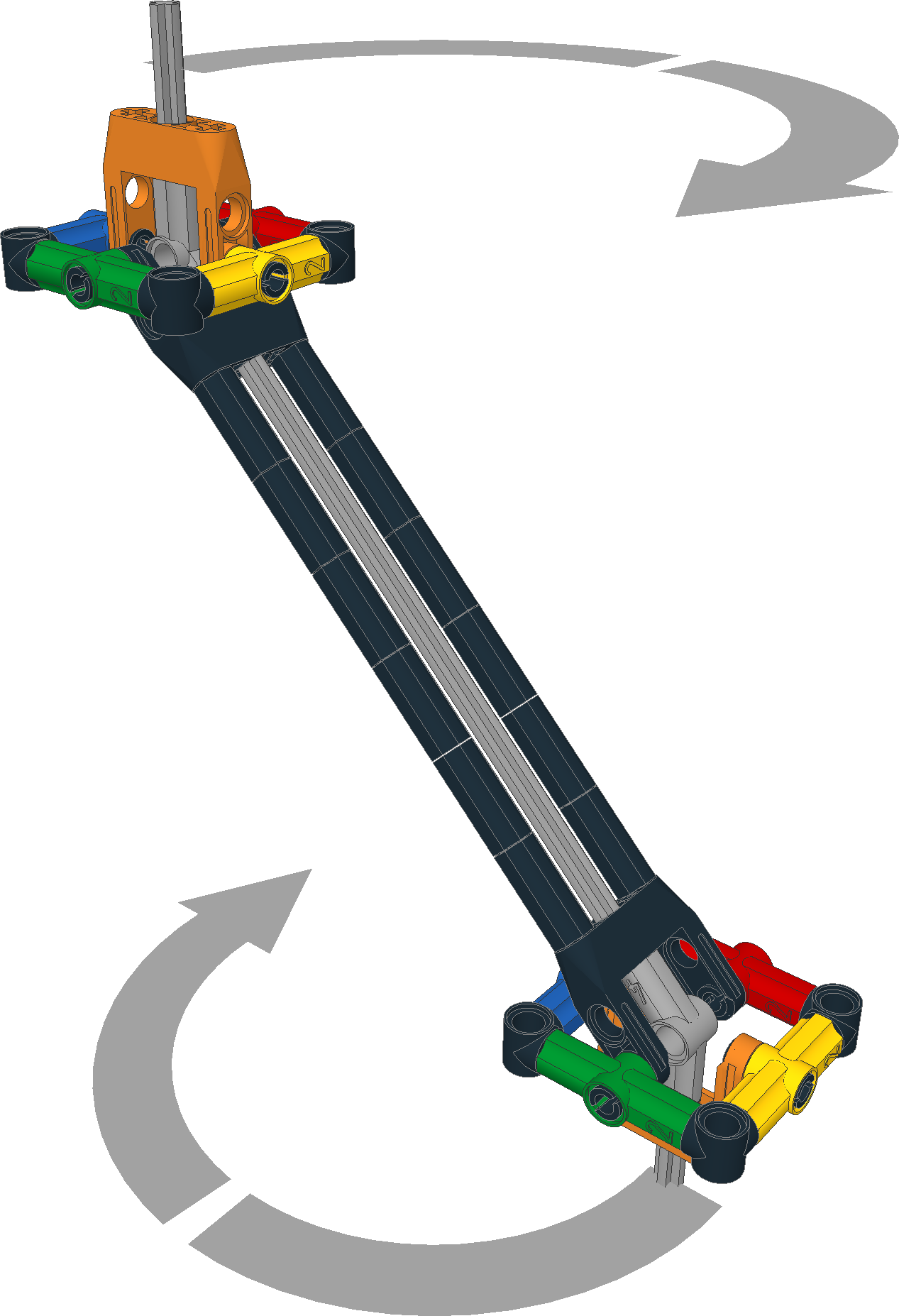}
  \caption{Parallel motion}
  \label{straight}
  \end{subfigure}
  \caption{Motion of the upper section.}
\end{figure}

If the core of the lower section is held fixed, the joints roll around the core like (half of) a smoke ring.  Hence the yokes at the two ends spin in opposite directions as shown in \cref{curve}.  If the core of the upper section is held fixed, the two end yokes spin in the same direction as shown in \cref{straight2}, maintaining the same orientation as each other in a similar manner to a parallelogram linkage.  Alternatively, we can hold the orientations of the two end yokes constant and swing them around each other as in \cref{straight}.

\begin{figure}
  \centering
  \includegraphics[width=.34\linewidth]{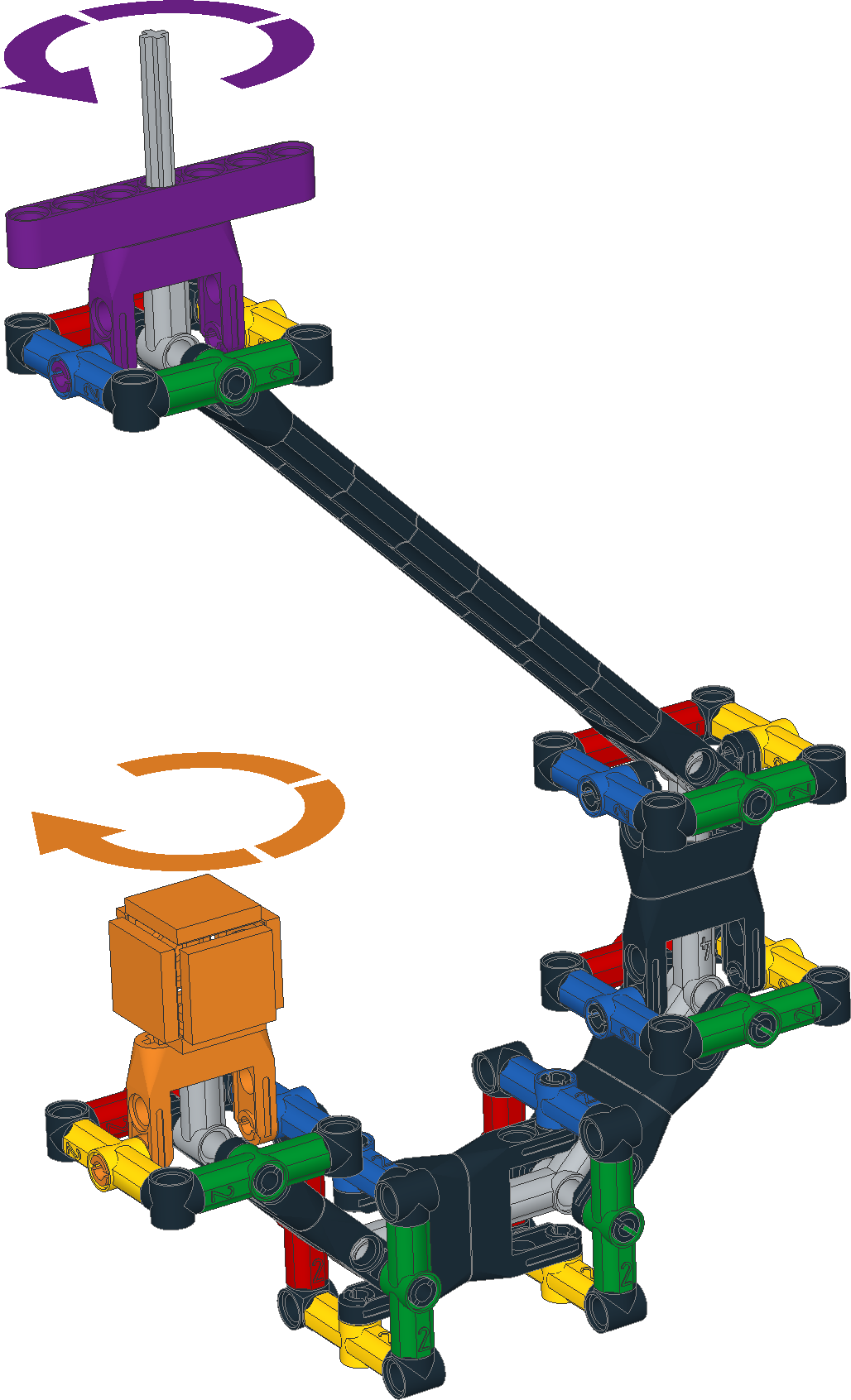}
  \caption{Motion with the core held fixed.}
  \label{fixed}
\end{figure}
The motion of the entire linkage results from combining the motions of the two sections.  If the core is held fixed, the two ends (which are attached to the frame and the cube) spin in opposite directions, at the same speed, as in \cref{fixed} -- also see the end of the video \cite{vid}.  Because of the constant-velocity property, both rotations are uniform.  Finally, spinning the entire structure at the same time results in the motion observed: the frame is stationary, while the cube spins at twice the control shaft's speed, in the same direction.  It is interesting that this $2:1$ ratio is achieved without gearing of any kind.

\section{Variants}

We mention several variations of possible expository and artistic merit.

\begin{enumerate}
\item As a purely visual embellishment, we can afix to each of the flat units of \cref{flat} a ring of the same shape and color scheme as those in \cref{rings}.  See \cref{bigger} for a LEGO implementation.  This makes the essential similarity of the 13 units more apparent, and highlights the pivoting relative motion of each consecutive pair.  On the other hand it becomes harder to see the underlying mechanism.
\begin{figure}
  \centering
  \includegraphics[width=.6\linewidth]{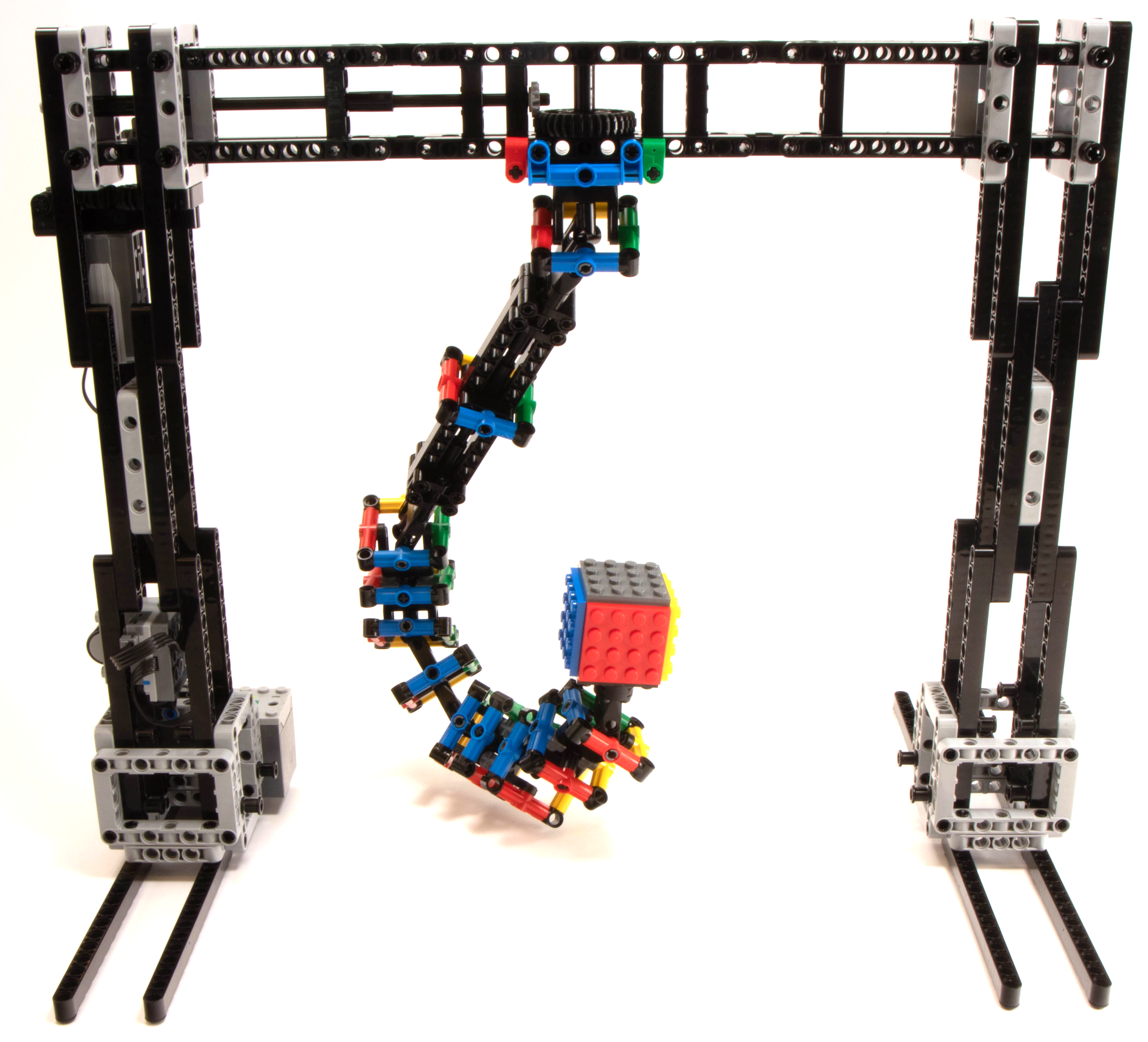}
  \caption{Embellishment with further rings.}
  \label{bigger}
\end{figure}

\item Two copies of the mechanism can be combined in an extended S-shape as in \cref{double}.  The core continues through a hole in the central cube as one rigid piece, emerging at the top and bottom.  The mechanism is controlled by rotating the core around a vertical axis while holding the top and bottom segments of the arm fixed.
\begin{figure}
  \centering
  \includegraphics[width=.45\linewidth]{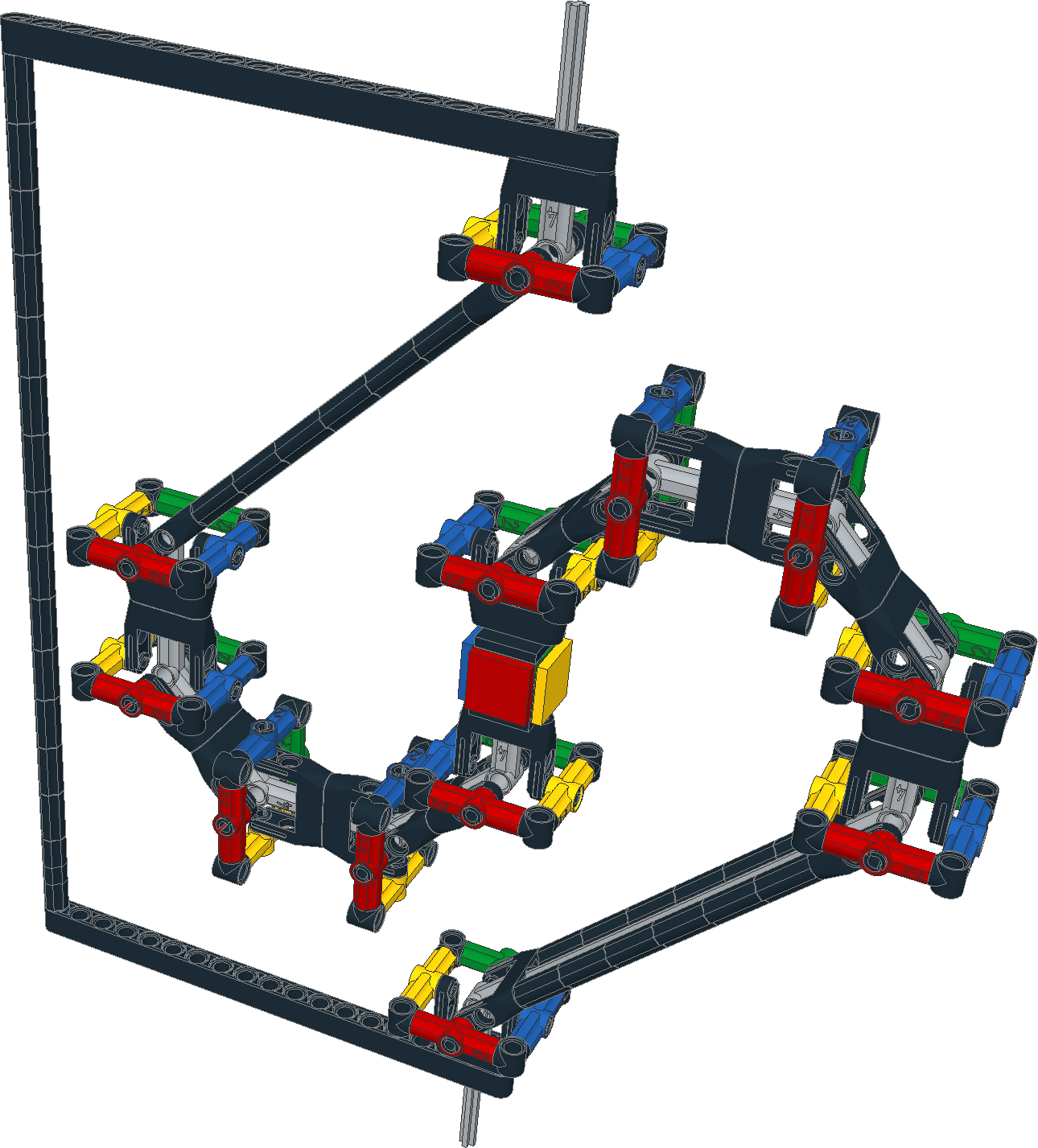}
  \caption{Double version.}
  \label{double}
\end{figure}

\item As indicated earlier, a string or cord can be attached along the outside of the arm, and tied off to the cube and the fixed frame at its two ends.  In order to minimize changes in length (which tend to lead to resistance or tangles), it is suggested to thread the string diagonally between counter-pivoting corners on the short sections, and parallel along the long straight section, as shown in \cref{string-detail}.  (This results in the cord spiralling around the curved section of the arm).
 \begin{figure}
  \centering
  \includegraphics[width=.3\linewidth]{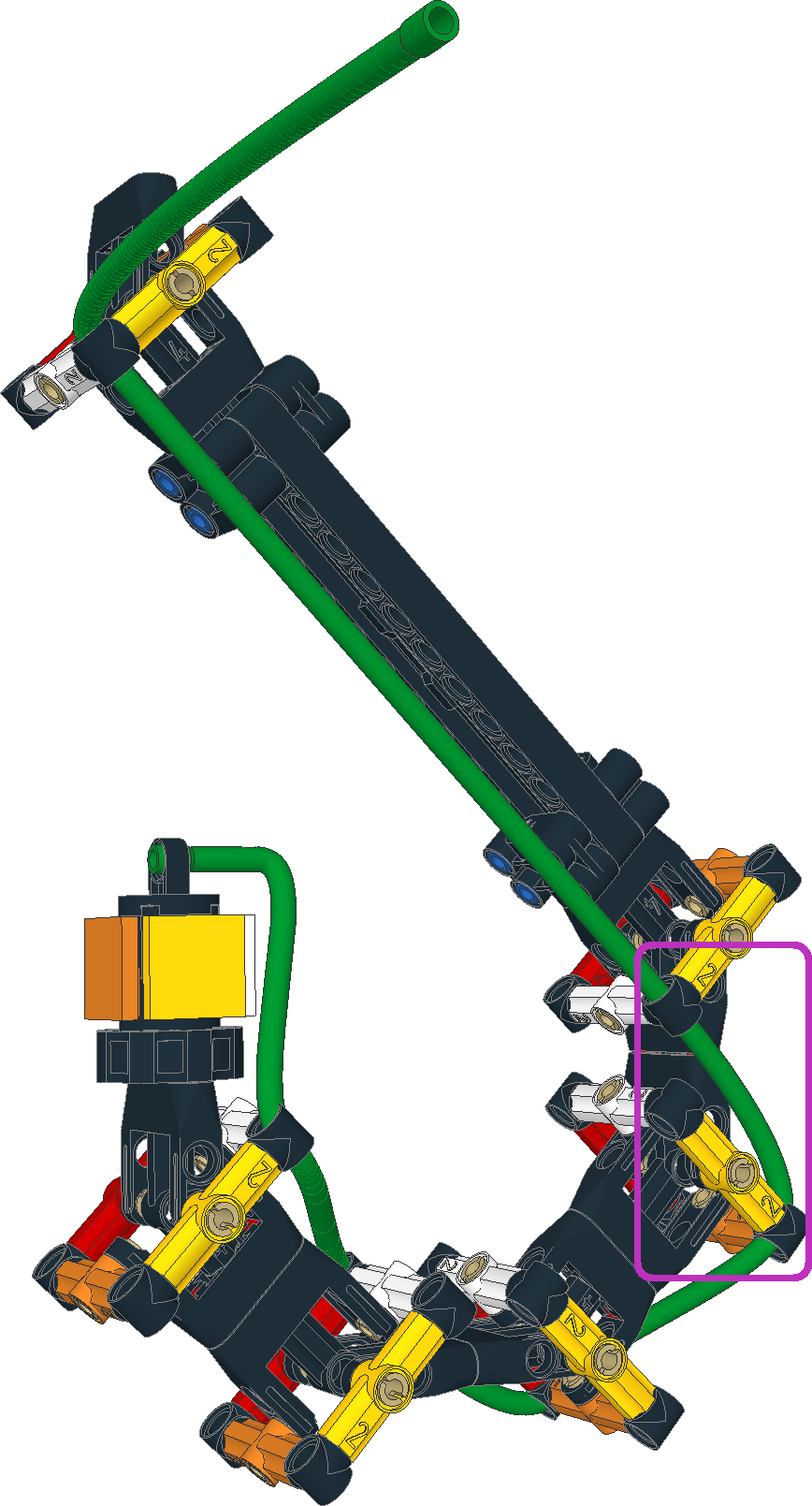}
  \caption{Attaching a string.}
  \label{string-detail}
\end{figure}

\item Extending the cord idea, since the control shaft is internal, the entire arm could be enclosed in a flexible tube running along its length, perhaps made of plastic, or perhaps crocheted or knitted.

\item The device can be used an anti-twist mechanism.  The cord described above can be replaced with wires or tubes, allowing electricity or fluid to be carried to a spinning object without the need for rotating contacts or seals.  Thus one could for instance add lights or water jets to the cube. The practical applications of such anti-twist mechanisms seem limited, because it is impossible to access the spinning object continuously from any fixed direction: the arm always gets in the way.  Adams' device \cite{adams} seems to suffer from the same difficulty.  It would be interesting to investigate this issue further.

\item Each consecutive pair of joints may be offset at a $90^\circ$ angle around the core, as shown in \cref{offset}.  Instead of producing the constant-velocity motion, this will have the opposite effect.  The non-uniformity resulting from universal joints will build up successively along the length of the arm.  When the control shaft is rotated uniformly, the cube should advance in sudden spurts (akin to the \emph{spotting} head movement technique employed by dancers).
\begin{figure}
  \centering
  \includegraphics[width=.45\linewidth]{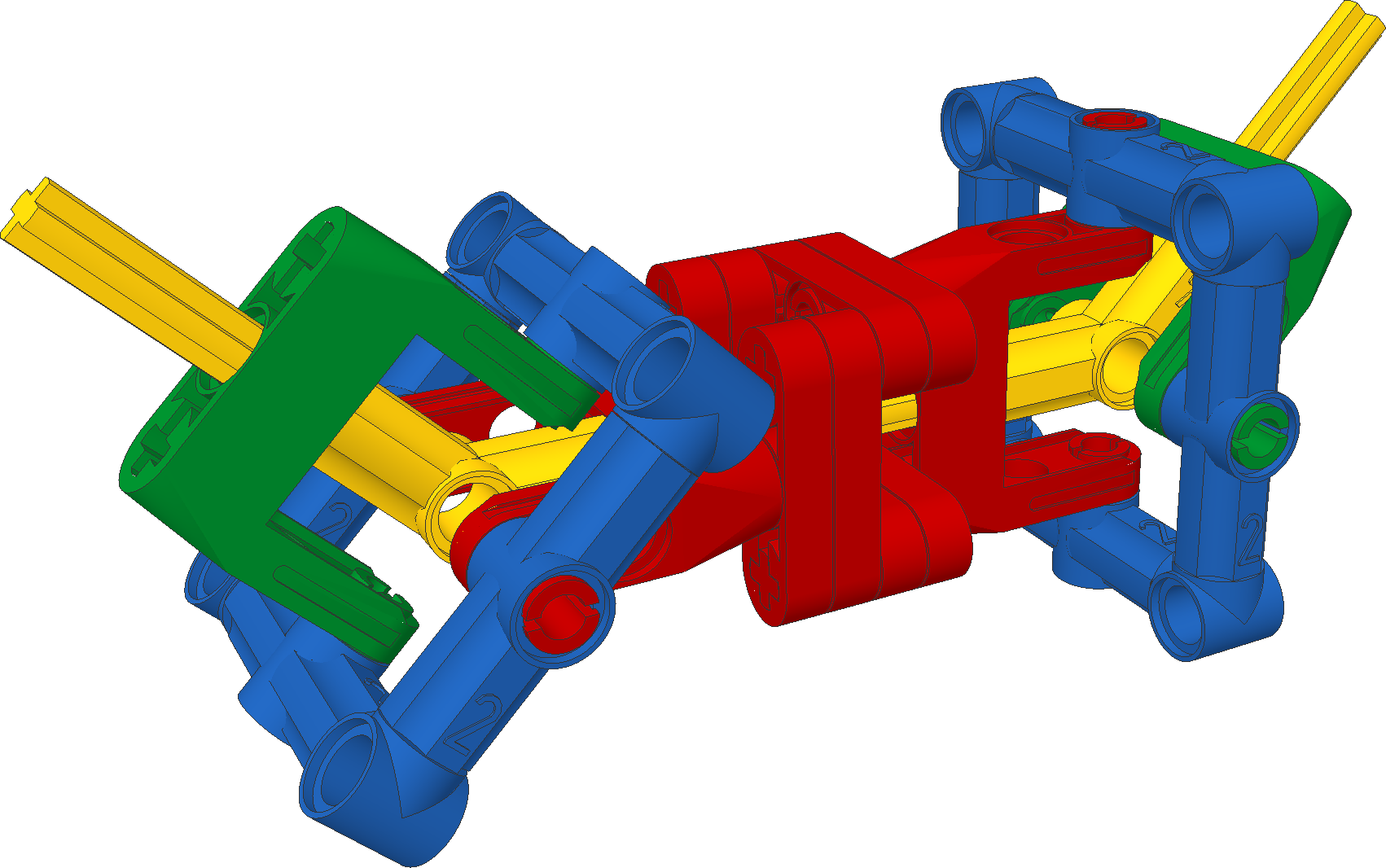}
  \includegraphics[width=.15\linewidth]{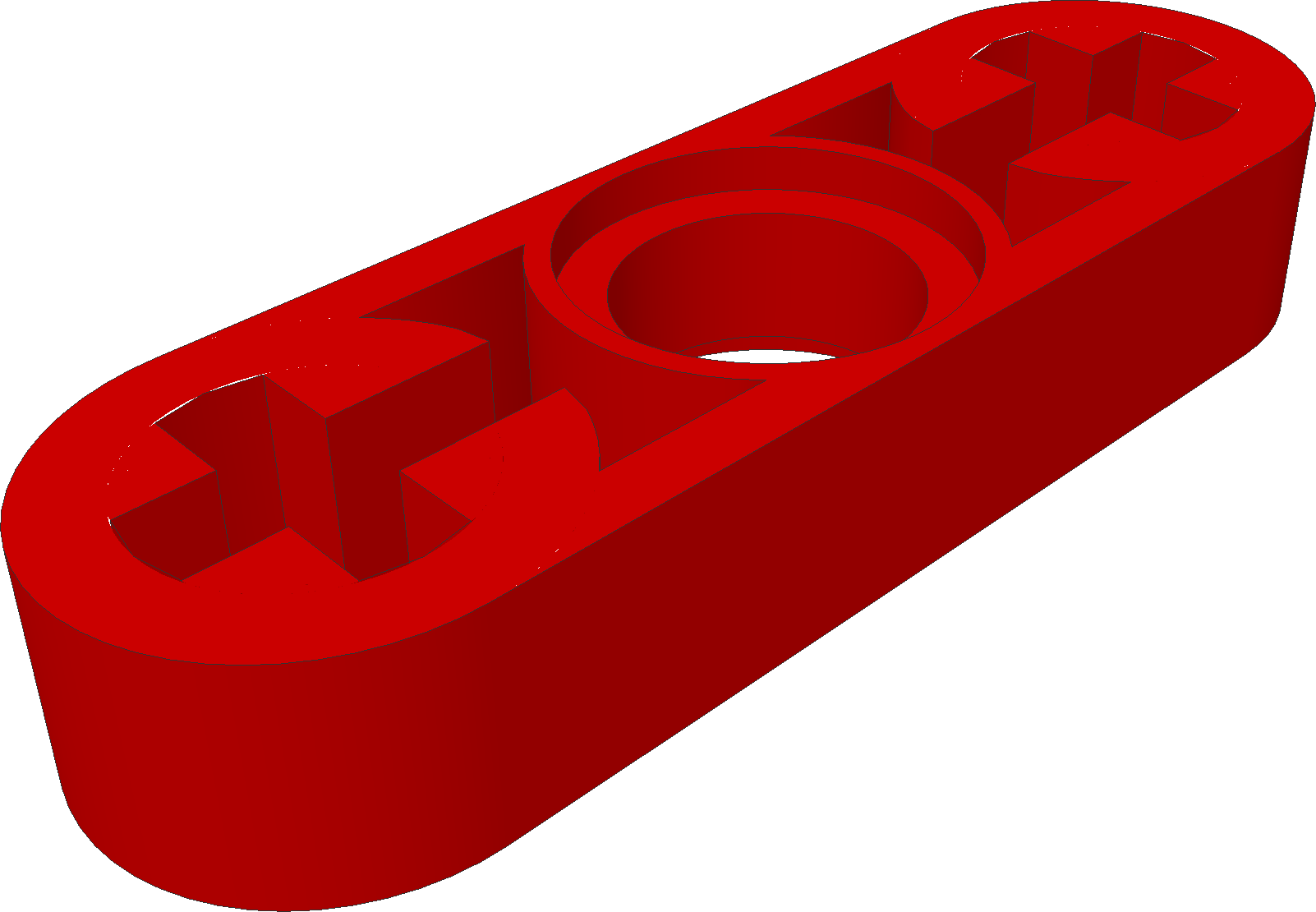}
  \caption{Offsetting the joints. (The relevant additional LEGO part is shown on the right.)}
  \label{offset}
\end{figure}
\end{enumerate}

See the YouTube channel of the author at \cite{vid} for versions of (1), (2), (3) and (5).

\section{Open Problems}

We suggest the following avenues for further investigation.

\begin{enumerate}
  \item
  We can generalize as follows.
  Consider a linkage consisting of a sequence of $n+1$ rigid bodies connected by $n$ hinge joints.  Suppose that the $n$th hinge joint is constrained to permit motion through a closed interval of angles of length $\alpha_i$. In our example, $n=12$ and $\alpha_i=\pi/2=90^\circ$ for each $i$.  For which $n$ and $\alpha_1,\ldots,\alpha_n$ does there exist such a linkage with the property that if the rigid body at one end is held fixed, then the orientation of the body at the other end can be varied continuously so as to execute endless uniform spinning around a fixed axis?

  There are many variations.  How does the answer change if the spinning is allowed to pause or go backwards during the motion, as long as it continues to advance in the long run?  How does the answer change if the end body is not allowed to undergo translation perpendicular to the axis around which it spins?  Similarly, what if translation along the axis is also forbidden (as is the case for our mechanism)?  Do any answers change if the rigid bodies are allowed to pass through each other?

  These questions are relevant to human performances of the plate trick or candle dance.  They can also be interpreted in the context of mechanical robot arms like the one proposed in this article, in which case control mechanisms are also of interest.

  Here is a purely algebraic formulation of the version of the question in which we permit arbitrary translations and pass-through, but the spinning is required to be monotone and without pauses.  Let $S=\{a+b\i+c\j+d\k:a^2+b^2+c^2+d^2=1\}$ be the sphere of unit quaternions.  Let $\rho(\theta)=\cos (\theta/2)+\i\sin(\theta/2)\in S$, which maps to a rotation by angle $\theta$ about a fixed axis under the double cover from $S$ to $SO(3)$.  For which $n$ and which $\alpha_1,\ldots,\alpha_n\in[0,4\pi]$ do there exist $\kappa_1,\ldots,\kappa_n\in S$ and continuous functions $\theta_i:[0,4\pi]\to[0,\alpha_i]$ satisfying $\theta_i(0)=\theta_i(4\pi)$ for each $i$, for which
  $$\rho(\theta_1(t))\kappa_1\cdots\rho(\theta_n(t))\kappa_n=\rho(t)\qquad\forall t\in [0,4\pi]?$$
  (Here, the $\kappa_i$ represent the relationships between successive hinge joint axes, and the $\theta_i$ represent their equations of motion.)

  More specifically, for each $n$, what is the minimum (or infimum)  $\alpha=\alpha(n)$ such that we can take $\alpha_i=\alpha$ for all $i$?  Clearly $\alpha(1)=4\pi$, while the mechanism of this article shows that $\alpha(12)\leq \pi/2$.
  
  Somewhat related questions are considered in \cite{waving}.

  \item
  Formulate and prove (or disprove) a precise version of ``the impracticality of anti-twist mechanisms''.  Roughly: if a continuously spinning object is tethered to the ground, then any ray emanating from the object and pointing in a fixed direction must at some point intersect the tether.

\begin{figure}
  \centering
  \includegraphics[width=.45\linewidth]{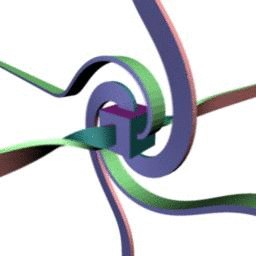}
  \caption{Belts in six directions; image by Jason Hise.}
  \label{cube}
\end{figure}
  \item
  A beautiful animation by Jason Hise (see \cref{cube} and \cite{hise}) shows a spinning cube tethered by belts attached to all six faces, rather than just the two in \cref{double}, with the belts neither twisting up nor intersecting one another.  (In fact, arbitrarily many belts in arbitrary directions are possible -- see Hise's YouTube channel \cite{hise} for more).  Can all six belts be emulated by mechanical linkages?
\end{enumerate}

\subsection*{Acknowledgements}

This project was conceived at the 2019 Illustrating Mathematics programme at The Institute for Computational and Experimental Research in Mathematics (ICERM), inspired by a beautiful lecture series by Richard Schwartz of Brown University.  I thank Richard Schwartz, ICERM, and the organizers.

\bibliographystyle{abbrv}
\bibliography{trunk}
\addresseshere

\newpage
\appendix
\section*{Appendix: the LEGO model}

We provide full instructions for constructing a working model of the spinor linkage from LEGO parts.  The list of required parts (with identification numbers) is given first, followed by pictorial building instructions.  These materials, and most of the images in the article, were generated using the rich ecosystem of free third-party applications generously written and maintained by the community of LEGO enthusiasts: principally LDraw, MLCad, LDView and LPub, and rebrickable.com for the parts list.

Various approaches are available for acquiring the LEGO parts, with trade-offs between cost and convenience.  The excellent specialist marketplaces \url{http://bricklink.com} and \url{http://brickowl.com} currently constitute the mainstay of the market for specific parts.  Applications and features are available for maintaining and fulfilling lists of parts, including partially automated tools for sourcing parts from multiple sellers.  New LEGO sets and second-hand lots can be economical routes to acquiring less targeted collections of parts.  There is plentiful advice online about the process.  Details are likely to evolve over time.

The website \url{http://rebrickable.com} is a highly recommended starting point linking many relevant resources.  The rebrickable page for this model, including the parts list, is essentially a one-stop solution \cite{rebrick}:

\url{http://rebrickable.com/mocs/MOC-50050/aeh5040/spinor-linkage/}

The colour scheme was chosen in order to balance elegance and clarity with availability of parts (which affects cost).  Of course, one can choose different colours.  In particular, availability of colours is likely to change over time -- it is worth investigating colour substitutions if costs seem unduly high.  One also has a choice between new and used parts, with a corresponding trade-off between quality and cost.

\newpage
{\centering
Parts list (generated using rebrickable.com)\\
\includegraphics[width=.95\textwidth]{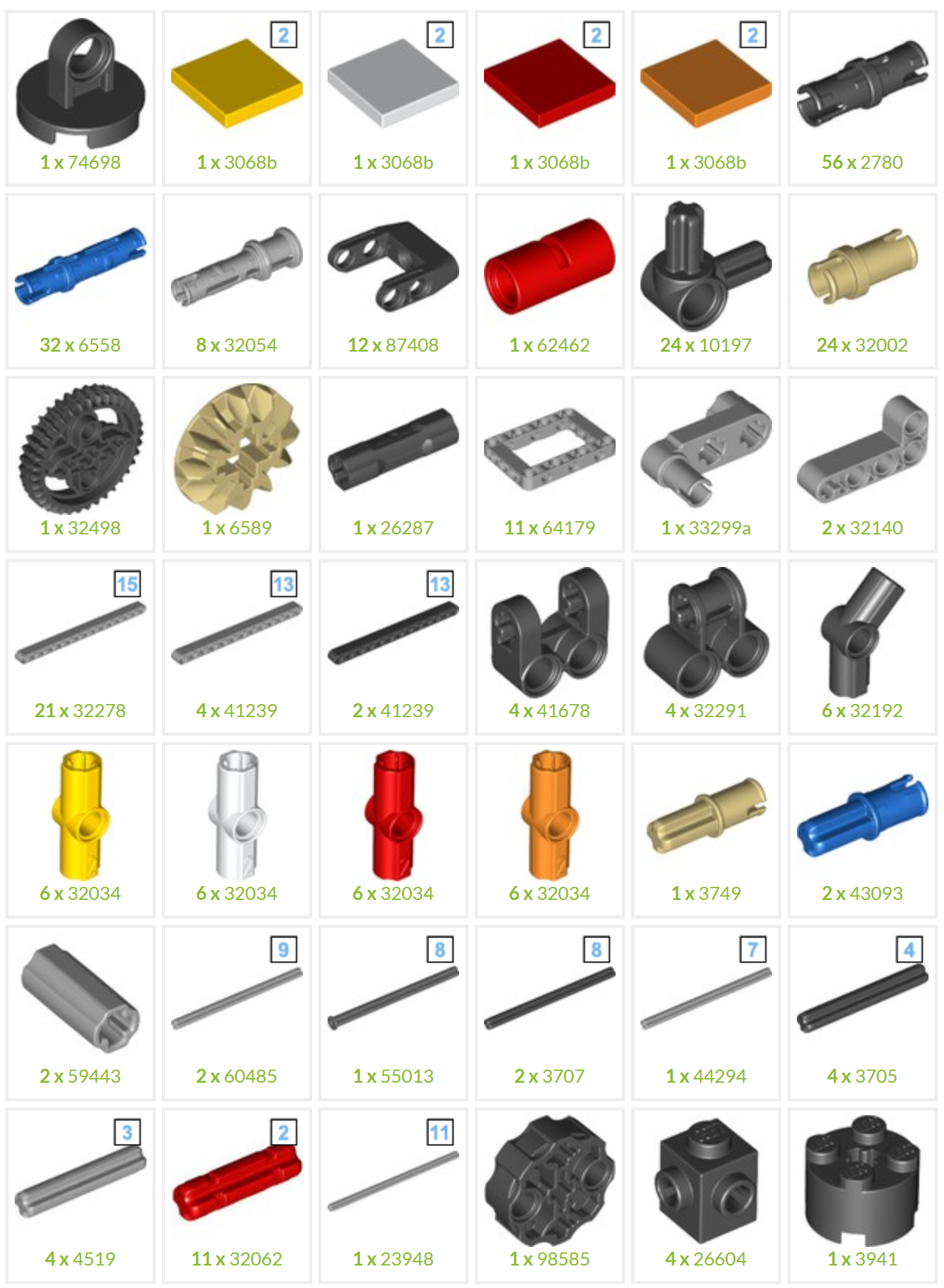}
}

\includepdf[pages=-]{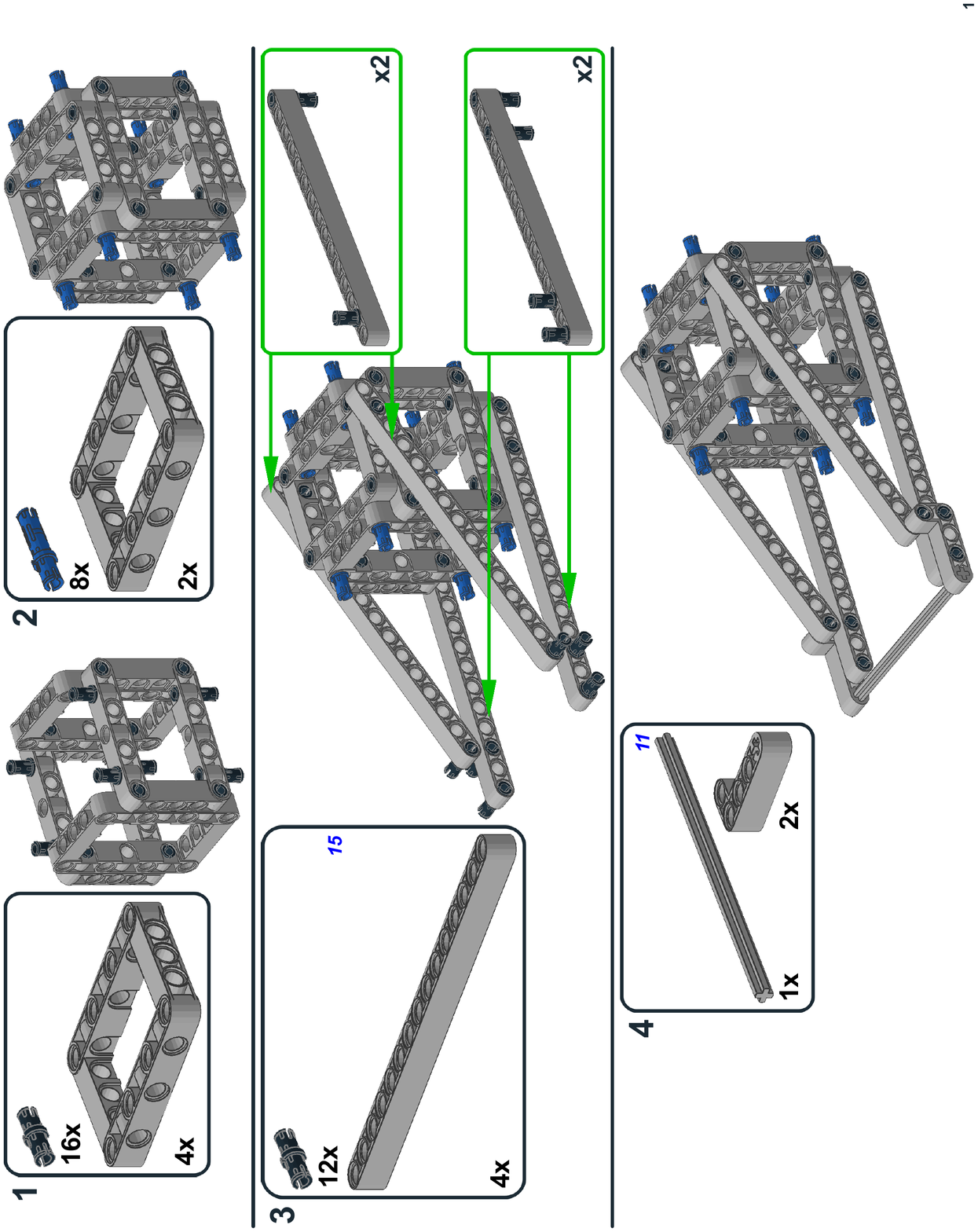}

\end{document}